\documentclass{aastex}
\usepackage{spr-astr-addons}
\usepackage{url}

\newcommand{\emaila}{khaibrakhmanovsa@gmail.com}
\newcommand{\lti}[1]{\mbox{{\footnotesize #1}}}

\begin{document}

\title{Fossil magnetic field of accretion disks of young stars}
\slugcomment{Not to appear in Nonlearned J., 45.}
\shorttitle{Short article title}
\shortauthors{Authors et al.}

\author{Dudorov A.~E.\altaffilmark{1}} \and \author{Khaibrakhmanov S.~A.\altaffilmark{1}}
\email{\emaila}

\altaffiltext{1}{Chelyabinsk state university, Chelyabinsk, Russia}

\begin{abstract}

We elaborate the model of accretion disks of young stars with the fossil large-scale magnetic field in the frame of Shakura and Sunyaev approximation. Equations of the MHD model include Shakura and Sunyaev equations, induction equation and equations of ionization balance. Magnetic field is determined taking into account ohmic diffusion, magnetic ambipolar diffusion and buoyancy. Ionization fraction is calculated considering ionization by cosmic rays and X-rays, thermal ionization, radiative recombinations and recombinations on the dust grains.
 
Analytical solution and numerical investigations show that the magnetic field is coupled to the gas in the case of radiative recombinations. Magnetic field is quasi-azimuthal close to accretion disk inner boundary and quasi-radial in the outer regions.

Magnetic field is quasi-poloidal in the dusty ``dead'' zones with low ionization degree, where ohmic diffusion is efficient. Magnetic ambipolar diffusion reduces vertical magnetic field in 10 times comparing to the frozen-in field in this region. Magnetic field is quasi-azimuthal close to the outer boundary of accretion disks for standard ionization rates and dust grain size $a_d=0.1\,\mu$m. In the case of large dust grains ($a_d>0.1\,\mu$m) or enhanced ionization rates, the magnetic field is quasi-radial in the outer regions.

It is shown that the inner boundary of dusty ``dead'' zone is placed at $r=(0.1-0.6)$ AU for accretion disks of stars with $M=(0.5-2)\,M_{\odot}$. Outer boundary of ``dead'' zone is placed at $r=(3-21)$ AU and it is determined by magnetic ambipolar diffusion. Mass of solid material in the ``dead'' zone is more than 3 $M_{\oplus}$ for stars with $M\geq 1\,M_{\odot}$.
\end{abstract}


\section{Introduction}

Observations indicate that stars born at present time in magnetized rotating cores of molecular clouds~\citep{shu87, dudorov91, dudorov95, mckee07}. Centrifugal and electromagnetic forces lead to the formation of a disk-like structure during the gravitational collapse of the molecular cloud cores. Evolution of accretion disks depends on efficiency of angular momentum removal. Turbulence, magnetic braking and outflows are the most important mechanisms of angular momentum transfer in accretion disks of young stars. Turbulence in accretion disks comes probably from magnetorotational instability \citep[MRI, ][]{balbus91}. Magnetic braking mechanism is based on the process of angular momentum transfer by torsional Alfven waves \citep{alfven_book}. Centrifugally driven winds arise when ordered magnetic field lines are inclined more than 30 degrees from vertical \citep{blandford82}. Efficiency of angular momentum transport depends on strength and geometry of the magnetic field.

Numerical simulations indicate that initial magnetic flux of molecular clouds cores is partially conserved during the process of star formation \citep[e.g.,][]{dud87}. \cite{dudorov08} shown that the initially uniform magnetic field acquires hour-glass geometry during protostellar cloud collapse. Collapsing protostellar cloud with magnetic field evolves into flat structure according to numerical simulations of \cite{dudorov03}. They shown that the collapse of the clouds with strong magnetic field switches to the magnetostatic contraction into oblate self-graviting structures. These numerical simulations show that accretion disks of young stars should have fossil magnetic field. In other words,  the magnetic flux of accretion disks is the relic of parent protostellar clouds magnetic flux.

There are some observational confirmations of predictions of the fossil magnetic field theory \citep{dudorov95}. \cite{girart06} found hour-glass shaped geometry of the magnetic field in the low-mass protostellar system NGC 1333 IRAS 4A using the Submillimeter Array polarimetry observations at 877 $\mu$m dust continuum emission. \cite{chapman13} found some evidence of correlation between core magnetic field direction and protostellar disk symmetry axis in several low-mass protostellar cores using the SHARP polarimeter at Caltech Submillimeter Observatory at 350 $\mu$m dust continuum emission. \cite{donati05} probably detected magnetic field with azimuthal component of order of 1 kGs in the FU Orionis accretion disk. Observational data are scanty, so theoretical investigation of magnetic field of accretion disks of young stars is the important problem.

Dragging of the magnetic field in the accretion disks has been investigated in several works. \cite{lubow94}, \cite{agapitou96} considered bending of the initially vertical weak magnetic field in infinitesimally thin viscous disk. They pointed out that inclination of poloidal magnetic fields lines and wind launching possibility depend on magnetic diffusion efficiency which is characterized by the magnetic Prandtl number $\mathcal{P}$. \cite{RRS96} showed that in the case of rotating disk, in addition to radial magnetic field component, strong toroidal component of the magnetic field is also generated when magnetic diffusion is feeble, $\mathcal{P}\gg 1$. \cite{shalybkov00} investigated steady-state structure of the accretion disk at the presence of dynamically important magnetic field. They found that sufficient inclination of the poloidal magnetic field lines to rotation axis is achieved also in the case of intermediate magnetic diffusion efficiency ($\mathcal{P}\simeq 1$). \cite{guilet12} pointed out that the accretion disk vertical structure has great effect on the magnetic flux evolution. In the limit of high conductivity, \cite{okuzumi13a} obtained analytical profile of the vertical component of the magnetic field, $B \propto r^{-2}$, that follows from magnetic flux conservation and gives upper estimation of the magnetic field strength in accretion disks, 0.1 Gs at 1 AU and $\sim$1 mGs at 10 AU.

Ohmic dissipation (OD), magnetic ambipolar diffusion (MAD), turbulent diffusion and buoyancy are the basic dissipation mechanisms limiting fossil magnetic field during protostellar cloud collapse and subsequent accretion \citep[e.g.,][]{dud87}. Efficiency of OD and MAD depends on the ionization fraction. Cosmic rays \citep{gammie96} and stellar X-rays \citep{igea97} ionize effectively only surface layers of accretion disks of young stars. Regions of low ionization fraction (``dead'' zones) arise near accretion disk midplane under such circumstances. Magnetic diffusion suppresses MRI in the ``dead'' zones. Turbulence attenuation in the ``dead'' zones favours matter accumulation, gravitational instability and planet formation \citep[e.g.,][]{armitage_book}. There are three important non-ideal magnetohydrodynamical (MHD) effects operating in accretion disks: OD, MAD and Hall effect \citep[e.g.,][]{wardle07}. Efficiency of MAD and Hall effect depends on magnetic field strength. Geometry of the magnetic field also plays crucial role in the producing of MRI-induced turbulence. \cite{simon13b} showed that the vertical magnetic field enhances turbulent stresses comparing to the case with the purely toroidal magnetic field.

Global numerical investigations of magnetized accretion disks and ``dead'' zones were performed in the ideal MHD limit \citep{fromang06} and in resistive limit \citep{dzyurkevich10} with magnetic field strength being free parameter. \cite{bai11}, \cite{simon13a, simon13b} performed calculations with MAD in the frame of the local shearing-box approximation with fixed magnetic field strength and/or geometry. 

Numerical modelling of magnetized accretion disks is still challenging problem, so semi-analytical models are widely used such as the minimum mass solar nebula model \citep[MMSN, e.g., ][]{weidenschiling77} and $\alpha$-model of \cite{ss73}. Investigations of ``dead'' zones in accretion disks in most cases use MMSN model \citep{sano00, bai09, mohanty13} or $\alpha$-model \citep{gammie96, fromang02, terquem08, martin12}. Large-scale magnetic field is ignored in these models. Prescribed magnetic field is also used in the models of magnetically driven winds \citep[e.g.,][]{CSD_book}. Recently, \cite{bai13a} and \cite{bai13b} performed numerical simulations in frame of local ``shearing-box'' approximation and showed that angular momentum transport is driven by magnetocentrifugal wind in the laminar ``dead'' zones, while MRI operates outside these regions.

There are several papers concerning computation of the magnetic field of accretion disks. \cite{dekool99} constructed accretion disk model taking into account MHD turbulence. In order to incorporate magnetic field into the model, they used proportionality of the magnetic stresses and the Reynolds stresses \citep{stone96}. \cite{martin12} used similar approach in order to determine ``dead'' zones boundaries. \cite{vorobyov06} estimated vertical magnetic field $B_z$ in the accretion disk assuming flux freezing and using the mass-to-flux ratio $\lambda=2\pi G^{1/2}\Sigma / B_z$ \citep{nakano78}, where $\Sigma$ -- accretion disk surface density. \cite{shu07} made analytical estimations of vertical component of magnetic field from the equation of centrifugal balance in protoplanetary disk diluted by poloidal magnetic tension. They neglected azimuthal magnetic field component.

We investigate the strength and geometry of the fossil large-scale magnetic field of accretion disks of young stars taking into account ohmic diffusion, magnetic ambipolar diffusion and buoyancy. The accretion disk model is based on the approximations of $\alpha$-model \citep{ss73}. We adopt kinematic approach and neglect magnetic field influence on the accretion disk structure. Ionization equations include shock ionization by X-rays, cosmic rays and radionuclides, thermal ionization, radiative recombinations and recombinations on the dust grains. Characteristics of ``dead'' zones in accretion disks of young stars are investigated in the frame of the elaborated model.

The paper is organized as follows. In section \ref{Sec:Model}, we formulate basic equations of our model. In section \ref{Sec:AnalytResults}, we obtain and analyse analytical solution of the model equations. Results of numerical calculations of the magnetized accretion disks structure are presented in the section \ref{Sec:NumResults}. In section \ref{Sec:DZ}, results of investigation of ``dead'' zones characteristics are presented. We summarize and discuss our results in section \ref{Sec:Discussion}.

\section{Accretion disk MHD model}
\label{Sec:Model}

\subsection{Basic equations}
In order to investigate the dynamics of accretion disks with large-scale magnetic field, we consider equations of magnetohydrodynamics \citep{LL_VI} taking into account ohmic and magnetic ambipolar diffusion \citep{dud87}
{
\allowdisplaybreaks
\begin{eqnarray}
  \frac{\partial\rho}{\partial t} + \mbox{div}\left(\rho\textbf{V}\right) &=& 0, \label{Eq:Cont}\\
  \rho\frac{\partial {\bf V}}{\partial t} + \left(\rho{\bf V}\nabla\right){\bf V} &=& -\nabla P - \rho{\bf \nabla} \Phi \nonumber\\
  & & + \mbox{div} \hat{\sigma}'\nonumber\\ 
  & & + \frac{1}{4\pi}\left[\mbox{rot}\,{\bf B}, {\bf B}\right],\label{Eq:Motion}\\
  \frac{\partial}{\partial t}\left(\rho\left(\varepsilon + \frac{V^2}{2} + \Phi\right) + \frac{B^2}{8\pi}\right) &=& -\mbox{div} {\bf F},\label{Eq:TotalEnergy}\\
  \frac{\partial\textbf{B}}{\partial t} &=& \mbox{rot}\left[\left(\textbf{V} + \textbf{V}_{\lti{ad}}\right), \textbf{B}\right] \nonumber\\
  & & -\mbox{rot}\left(\nu_{m} \mbox{rot}\textbf{B}\right), \label{Eq:Induction}
\end{eqnarray}}
where $\rho$, ${\bf V}$ $P$, $\varepsilon$ -- density, velocity, pressure and internal energy density of the gas, $\hat{\sigma}'$ -- viscous stress tensor, $\Phi$ -- gravitational potential, ${\bf B}$ -- magnetic induction vector, ${\bf V}_{\lti{ad}}$ -- ambipolar diffusion velocity, $\nu_{\lti{m}}$ -- ohmic diffusion coefficient, ${\bf F}$ -- energy flux density,
\begin{eqnarray}
	{\bf F} &=& \rho{\bf V}\left(\varepsilon + \frac{V^2}{2} + \Phi\right) - \left(V\hat{\sigma}'\right) + {\bf F}^{\lti{rad}}\nonumber\\
	& & + \frac{1}{4\pi}\left[{\bf B}, [{\bf V}, {\bf B}]\right] + \frac{c^2}{16\pi^2\sigma}\left[\mbox{rot}\,{\bf B}, {\bf B}\right],
\end{eqnarray}
${\bf F}^{\lti{rad}}$ -- radiation energy flux density. Coulomb conductivity $\sigma$ of the partially ionized plasma \citep{pikelner_book}
\begin{equation}
	\sigma = 1.5\times 10^{17} xT^{-1/2}\,\mbox{s}^{-1},\label{Eq:conductivity}
\end{equation}
where $T$ -- gas temperature, $x$ -- ionization fraction. Magnetic viscosity is
\begin{equation}
	\nu_{m} = \frac{c^2}{4\pi\sigma} = 480\,x^{-1}T^{1/2}\,\,\mbox{cm}^2\mbox{s}^{-1}.\label{Eq:ohmic}
\end{equation}

Magnetic ambipolar diffusion is the joint drift of charged particles through neutral gas under the action of electromagnetic force (MAD hereafter). Stationary MAD velocity \citep[e.g.,][]{dud87}
\begin{equation}
	{\bf V}_{\lti{ad}} = \frac{\left[\mbox{rot}\,{\bf B}, {\bf B}\right]}{4\pi x\rho^2\eta_{\lti{in}}},\label{Eq:MADvelocity}
\end{equation}
where
\begin{equation}
\eta_{\lti{in}} = \frac{m_{\lti{i}}\langle\sigma V\rangle_{\lti{in}}}{m_{\lti{n}}(m_{\lti{i}} + m_{\lti{n}})},
\end{equation}
-- ion-neutral interaction coefficient, neutral particle mass $m_{\lti{n}} = 2.3\, m_{H}$, mean ion mass $m_{\lti{i}} = 30\, m_{H}$, $m_{H}$ -- hydrogen atom mass, ion-neutral collision coefficient $\langle\sigma V\rangle_{\lti{in}}= 2.0\times 10^{-9}\,\mbox{cm}^3\,\mbox{s}^{-1}$ \citep{osterbrock61}.

\subsection{Kinematic MHD model of accretion disk}

In the kinematic approximation we neglect the electromagnetic force in the equation of motion comparing to the gravity force. In cylindrical coordinates system $\left(r,\,\varphi,\,z\right)$, velocity components ${\bf V} = \left(V_r,\,V_{\varphi},\,V_z\right)$ and magnetic induction components ${\bf B} = \left(B_r,\,B_{\varphi},\,B_z\right)$.  Self-gravity of the disk is neglected. Gravitational potential of the star with mass $M$
\begin{equation}
	\Phi = -\frac{GM}{\sqrt{r^2 + z^2}}.
\end{equation}
We assume that star has dipole magnetic field
\begin{equation}
	B_{\star} = B_{s}\left(\frac{R_s}{r}\right)^3,
\end{equation}
where $B_s$ is the the magnetic induction on stellar surface, $R_s$ -- stellar radius.

Equations of continuity (\ref{Eq:Cont}), motion (\ref{Eq:Motion}) and energy (\ref{Eq:TotalEnergy}) in axisymmetric case ($\partial /\partial \varphi = 0$) are reduced to
\begin{eqnarray}
	\frac{\partial \rho}{\partial t} + \frac{1}{r}\frac{\partial}{\partial r}\left(r\rho V_r\right) + \frac{\partial \rho V_z}{\partial z} &=& 0,\label{Eq:ContCyl}\\
	\rho\left(\frac{\partial V_r}{\partial t} + \left({\bf V}\nabla\right)V_r - \frac{V_{\varphi}^2}{r}\right) &=& -\frac{\partial P}{\partial r} - \rho\frac{\partial \Phi}{\partial r},\label{Eq:MotionR}\\
	\rho\left(\frac{\partial V_{\varphi}}{\partial t} + \left({\bf V}\nabla\right)V_{\varphi} + \frac{V_{\varphi}V_r}{r}\right) &=& \frac{1}{r}\frac{\partial}{\partial r}\left(r\sigma_{r\varphi}^{'}\right)\nonumber\\
	& & + \frac{\sigma_{r\varphi}^{'}}{r},\label{Eq:MotionPhi}\\
	\rho\left(\frac{\partial V_z}{\partial t} + \left({\bf V}\nabla\right)V_z\right) &=& -\frac{\partial P}{\partial z} - \rho\frac{\partial \Phi}{\partial z},\label{Eq:MotionZ}\\
	\frac{\partial}{\partial t}\left(\rho\left(\varepsilon + \frac{V^2}{2} + \Phi\right)\right) &=&  -\frac{1}{r}\frac{\partial}{\partial r}\left(rF_r\right)\nonumber\\
	& & - \frac{\partial F_{z}^{\lti{rad}}}{\partial z},\label{Eq:EnergyCyl}
\end{eqnarray}
where operator
\begin{equation}
	\left({\bf V}\nabla\right)V_i = V_r\frac{\partial V_i}{\partial r} + V_z\frac{\partial V_i}{\partial z}.
\end{equation}
Radial energy flux density
\begin{equation}
 F_{r} = \rho V_r\left(\varepsilon + \frac{V^2}{2} + \Phi\right) + \sigma_{r\varphi}^{'}V_{\varphi}.\label{Eq:Fr}
\end{equation}
Toroidal stress tensor component
\begin{equation}
	\sigma^{'}_{r\varphi} = \eta_{\lti{v}}r\frac{\partial \Omega}{\partial r},\label{Eq:StressTensor}
\end{equation}
where $\eta_{\lti{v}}$ -- dynamical viscosity coefficient, $\Omega = rV_{\varphi}$ -- angular velocity.

Induction equation (\ref{Eq:Induction}) can be written in the following way, after substituting (\ref{Eq:MADvelocity}) into it,
\begin{equation}
  \frac{\partial\textbf{B}}{\partial t} = \mbox{rot}\left[\textbf{V}, \textbf{B}\right] - \mbox{rot}\left(\eta \mbox{rot}\textbf{B}\right), \label{Eq:Induction2}
\end{equation}
where effective magnetic diffusion coefficient is introduced
\begin{equation}
	\eta = \nu_m + \eta_{\lti{ad}},\label{Eq:eta}
\end{equation}
and
\begin{equation}
	\eta_{\lti{ad}} = \frac{B^2}{4\pi x\rho^2\eta_{\lti{in}}}\label{Eq:etaMAD}
\end{equation}
-- MAD coefficient.

Induction equation (\ref{Eq:Induction2}) in cylindrical coordinates has the following components
{
\allowdisplaybreaks
\begin{eqnarray}
  \frac{\partial B_r}{\partial t} &=&  \frac{\partial \left(V_{r}B_{z}\right)}{\partial z} + \eta\frac{\partial^{2}B_{r}}{\partial z^{2}}\nonumber \\
   & & + \eta\left(\frac{1}{r}\frac{\partial}{\partial r}r\frac{\partial B_{r}}{\partial r} -
  \frac{B_{r}}{r^{2}}\right),\label{Eq:Br}\\
  \frac{\partial B_{\varphi}}{\partial t} &=& \frac{\partial \left(V_{r}B_{\varphi} - V_{\varphi}B_{r}\right)}{\partial r} +
  \frac{\partial \left(B_{z}V_{\varphi}\right)}{\partial z} + \eta\frac{\partial^{2}B_{\varphi}}{\partial z^{2}}\nonumber\\
  & & + \eta\left(\frac{1}{r}\frac{\partial}{\partial r}r\frac{\partial B_{\varphi}}{\partial r} -
  \frac{B_{\varphi}}{r^{2}}\right),\label{Eq:Bphi}\\
  \frac{\partial B_z}{\partial t} &=& -\frac{1}{r}\frac{\partial \left(rV_{r}B_{z}\right)}{\partial r} + \eta\frac{\partial^{2}B_{z}}{\partial z^{2}} \nonumber\\
  & & + \eta\left(\frac{1}{r}\frac{\partial}{\partial r}r\frac{\partial B_{z}}{\partial r}
  \right). \label{Eq:Bz}
\end{eqnarray}}

\subsubsection{Accretion disk structure}
\label{sec:ads}

We consider geometrically thin ($z \ll r$), optically thick accretion disk in steady state ($\partial/\partial t = 0$). Velocity components in the accretion disk satisfy the relations: $V_z\ll V_r \ll V_{\varphi}$. Convective terms in equations (\ref{Eq:MotionR}-\ref{Eq:MotionZ}) are negligible. Therefore and dimensional splitting of these equations can be realized. We solve equations of radial and vertical structure of accretion disk independently. 

Continuity equation (\ref{Eq:ContCyl}) expresses conservation of radial mass flux
\begin{equation}
	rV_r\rho = \mbox{const}.\label{Eq:ContStat}
\end{equation}
In the stationary case, equation (\ref{Eq:MotionR}) reduces to balance of centrifugal force and the radial component of the gravity force implying that the disk rotates with the angular velocity
\begin{equation}
	\Omega = \sqrt{\frac{GM}{r^3}}\left(1+\frac{z^2}{r^2}\right)^{-3/4}\label{Eq:Kepler},
\end{equation}
where $\Omega_{k} = \sqrt{GM/r^3}$ -- Keplerian angular velocity.

Equation (\ref{Eq:MotionZ}) transforms to the hydrostatic equilibrium equation that has the solution
\begin{equation}
	\rho(r,\,z) = \rho(r,\,0)\exp\left(-\frac{z^2}{2H^2}\right),\label{Eq:HydroStat}
\end{equation}
where $H$ is the accretion disk scale height,
\begin{equation}
	H = \frac{V_{\lti{s}}}{\Omega_k}.\label{Eq:H}
\end{equation}
We use equation of state $P=\rho V_s^2$ in equations (\ref{Eq:HydroStat}-\ref{Eq:H}), where $V_s = \sqrt{R_{g}T/\mu}$ is the isothermal sound velocity, $R_g$ -- universal gas constant.

Specific angular momentum $\rho\Omega^2r$ transfer in the accretion disk is described by $\varphi$-component of the equation of motion (\ref{Eq:MotionPhi}) that reduces to
\begin{equation}
	\frac{1}{r}\frac{\partial}{\partial r}\left(rV_r\rho\Omega^2 - r^2\sigma'_{r\varphi}\right) = 0.\label{Eq:AngMomStat}
\end{equation}
In frame of \citeauthor{ss73} model, molecular viscosity coefficient in (\ref{Eq:StressTensor}) is replaced by turbulent viscosity that is defined as
\begin{equation}
	\eta_{\lti{t}} = \alpha \rho V_{\lti{s}}H,
\end{equation}
where turbulence parameter $\alpha = V_{t} / V_{s}$, $V_t$ -- turbulent velocity.

Radial angular momentum transport by turbulent stresses leads to corresponding energy redistribution. 
Substituting radiation energy flux in diffusion approximation 

\begin{equation}
	F^{\lti{rad}}_z = -\frac{4}{3\kappa\rho}\frac{d\sigma_{\lti{sb}}T^4}{dz}
\end{equation}
into the energy equation (\ref{Eq:EnergyCyl}) together with (\ref{Eq:Fr}), we get
\begin{eqnarray}
	\frac{1}{r}\frac{\partial}{\partial r}\left(r\rho V_r\left(\frac{V_{\varphi}^2}{2} + \Phi\right) + \Omega r^2\sigma'_{r\varphi}\right) &=& \nonumber\\
	\frac{d}{dz}\left(\frac{4}{3\kappa\rho}\frac{d\sigma_{\lti{sb}}T^4}{dz}\right). & & \label{Eq:EnergyStat}
\end{eqnarray}
where $\kappa$ is the opacity coefficient, $\sigma_{\lti{sb}}$ -- Stephan-Boltzmann constant.

Turbulent stresses must vanish at the accretion disk inner boundary. We derive following equations from (\ref{Eq:ContStat}, \ref{Eq:AngMomStat}, \ref{Eq:EnergyStat}) in this case
\begin{eqnarray}
  \dot{M} &=& - 2 \pi r V_r \Sigma, \label{Eq:Mass}\\
  \dot{M} \Omega f &=& 2 \pi \alpha V_{\lti{s}}^2 \Sigma, \label{Eq:AngMom}\\
  \frac{4\sigma_{\lti{sb}} T^4}{3\kappa\Sigma} &=& \frac{3}{8\pi}\dot{M}\Omega^2 f,\label{Eq:Energy}
\end{eqnarray}
Equations (\ref{Eq:Mass}-\ref{Eq:Energy}) supplemented by (\ref{Eq:Kepler}) and (\ref{Eq:H}) represent standard equations system of \citeauthor{ss73} $\alpha$-model. Equation (\ref{Eq:Mass}) is derived from (\ref{Eq:ContStat}) with the help of integration over accretion disk scale height $H$. This equation determines relation between accretion rate $\dot M$ and surface density $\Sigma = 2 \left\langle \rho \right\rangle H$, where $\left\langle \rho \right\rangle$ -- {$z$-averaged} density. Equation (\ref{Eq:AngMom}) is the angular momentum conservation equation integrated over $H$, where $f \equiv 1 - \sqrt{r_{\lti{in}} / r}$, $r_{\lti{in}}$ -- accretion disk inner boundary. Energy equation (\ref{Eq:Energy}) is valid in the optically thick accretion disk. Energy spectrum of the accretion disk is characterized by effective temperature determined from the equation
\begin{equation}
	\sigma_{\lti{sb}}T_{\lti{eff}}^4 = \frac{3}{8\pi}\dot{M}\Omega^2 f,\label{Eq:Teff}
\end{equation}
reflecting balance between rate of viscous heating per unit of the accretion disk surface (right-hand side) and black body radiation energy flux density (left-hand side).

Note that equations (\ref{Eq:H}, \ref{Eq:Mass}) and (\ref{Eq:AngMom}) imply that
\begin{equation}
V_r = \alpha \left(\frac{H}{r}\right)^2V_{\varphi},\label{Eq:VrVphi}
\end{equation}
i.e. $V_r \ll V_{\varphi}$ in geometrically thin disk, since $\alpha<1$, $H\ll r$.

We specify opacity coefficient
\begin{equation}
\kappa = \kappa_0 \rho^{a}T^{b},
\end{equation}
in order to close equations system (\ref{Eq:Kepler}, \ref{Eq:H}, \ref{Eq:Mass}-\ref{Eq:Energy}) describing radial accretion disk structure. Parameters $a$, $b$, $\kappa_0$ are evaluated according to \citet{semenov03}. These constants are listed in the table \ref{Tab:1}. Dust grains consisting of ices, organic material, iron and silicates are the dominant opacity agents at the low temperatures ($T<2000$ K, rows 1-5 in the table~\ref{Tab:1}). Adopted dust composition is in qualitative agreement with composition of recovered fragments of the Chelyabinsk meteorite \citep{cb13}. Opacity coefficient does not depend on density in this temperature range. Dust is evaporated at 2000 K and molecular opacities become dominant up to temperatures of molecules dissociation $T=3000$ K (row 7 in the table~\ref{Tab:1}). Ionized gas opacities dominate at higher temperatures (rows 8-9 in the table \ref{Tab:1}).

\begin{table*}
\small
\caption{Opacity $\kappa=\kappa_0 \rho^a T^b$.}
\centering
\begin{tabular}{@{}cccccl@{}}
\tableline
region number & $\kappa_0$ & $a$ & $b$ & main opacity agent & temperature range\\
\tableline
1 & $2.0\times 10^{-4}$ & 0 & 2 & ice & $T<150$ K\\
2 & $1.4\times 10^{11}$ & 0 & -5 & ice evaporation & \\
3 & $0.01$ & 0 & 1.11 & organics & $T\in[170,\,\sim 400]$ K\\
4 & $3.16\times 10^{-16}$ & 0 & -6 & organics evaporation & \\
5 & $3.0\times 10^{-3}$ & 0 & 1 & iron, silicates & $T\in[\sim 450,\,\sim 1500]$ K\\
6 & $2.15\times 10^{34}$ & $1/3$ & -10 & dust evaporation & $T\in[\sim 1500,\,\sim 2000]$ K\\
7 & $0.002$ & 0 & 0 & molecules & $T\in[\sim 2000,\,\sim 3000]$ K\\
8 & $10^{-36}$ & $1/3$ & 10 & $H$-scattering & $T\in[\sim 3000,\,\sim 10^{4}]$ K\\
9 & $1.5\times 10^{20}$ & 1 & $-5/2$ & bound-free, free-free & $T > 10^4$ K\\
\tableline
\end{tabular}
\label{Tab:1}
\end{table*}

Accretion disks with $\dot M$ lower than few $10^{-8}\,M_{\odot}/\mbox{yr}$ are passive in terms of heating mechanism \citep[e.g., see][]{armitage_book}. Reprocessing of stellar radiation at large distances from the star must be included in the model. We adopt simple analytical profile of irradiation temperature \citep{hayashi81}
\begin{equation}
 T_{\lti{irr}} = 280\,\left(\frac{L_{\star}}{L_{\odot}}\right)^{1/4} \left(\frac{r}{1\,\mbox{AU}}\right)^{-1/2}\,\mbox{K},\label{Eq:TirrAppr}
\end{equation}
where $L_{\star}$ is luminosity of the star. 

\subsubsection{Ionization fraction}

Magnetic diffusion efficiency depends on the ionization fraction (see equations (\ref{Eq:conductivity}-\ref{Eq:MADvelocity})). We adopt ionization model from \cite{dud87}. Ionization fraction at low temperatures is calculated from the equation of ionization balance taking into account radiative recombinations and recombinations on the dust grains. We considered that ionization fraction at low temperatures is determined mainly by hydrogen collisional ionization,
\begin{equation}
 (1-x_{s})\xi = \alpha_r x_{s}^2 n + \alpha_{g} x_s n, \label{Eq:Ion}
\end{equation}
where $x_s = n_{e}/n$ -- ionization fraction, $n$ -- gas number density, $\xi$ -- ionization rate. Equation (\ref{Eq:Ion}) implies that ionization fraction of all species is equal to hydrogen ionization fraction. Radiative recombinations coefficient according to \cite{spitzer_book}
\begin{equation}
 \alpha_r = 2.07\times 10^{-11}T^{-1/2}\Phi(T),\, \mbox{cm}^3\mbox{s}^{-1}.
\end{equation}
Numerical factor $\Phi(T) = 3.0$ in temperature range ${10\,\mathrm{K}\,\leq T \leq 10^3\,\mathrm{K}}$ and $\Phi(T) = 1.5$ in temperature range $10^3$ K $\leq T \leq 10^4$ K. Coefficient of recombinations on dust grains,
\begin{equation}
\alpha_g = X_{\lti{g}}\langle\sigma_{\lti{g}}V_{\lti{ig}}\rangle,\label{Eq:Alpha_g}
\end{equation}
where $X_{\lti{g}}=n_{\lti{g}}/n_{\lti{H}}$ -- relative abundance of the dust grains, $\sigma_{\lti{g}}$ -- grain cross-section, $V_{\lti{ig}}$ -- ion-grains relative velocity. We adopt dust-to-gas mass ratio $Y_{\lti{g}}=0.01$ and fiducial dust grains size $a_d=0.1\,\mu$m.  Piecewise linear approximation of the temperature dependence of coefficient $\alpha_g$ is used in order to take into account dust grains evaporation. Recombination coefficient $\alpha_{\lti{g}} = \alpha_{\lti{g0}}=4.5\times 10^{-17}\,\mbox{cm}^3\mbox{s}^{-1}$ at temperatures less than ice evaporation temperature $T=150$ K. It linearly decreases with temperature down to the value $\alpha_{\lti{gm}}=3.0\times 10^{-18}\,\mbox{cm}^3\mbox{s}^{-1}$ in range $T\in \left[150,\,400\right]$ K. Then $\alpha_{\lti{g}}=\alpha_{\lti{gm}}$ right up to $T=1500$ K, and $\alpha_{\lti{g}}$ linearly decreases with temperature from $\alpha_{\lti{gm}}$ down to 0 in temperature range $T\in \left[1500,\,2000\right]$ K.

We do not take into account dissociative recombinations. We focus on the calculation of the intensity and geometry of the magnetic field in the limiting cases of high and low ionization fraction. Recombinations on dust grains determine minimal ionization fraction, and radiative recombinations determine maximal ionization fraction. Dissociative recombinations correspond to the intermediate case, so this mechanism will not change qualitative picture of our results.

Equation of ionization balance (\ref{Eq:Ion}) has two asymptotics. In the radiative recombinations case, $\alpha_{\lti{g}}=0$,
\begin{equation}
 x_{r} = \sqrt{\frac{\xi}{\alpha_r n}} \propto n^{-1/2},\label{Eq:XsRad}
\end{equation}
i.e. ionization fraction is inversely proportional to the square root of number density. In the case of recombinations on dust grains, $\alpha_r=0$,
\begin{equation}
 x_{d} = \frac{\xi}{\alpha_g n} \propto n^{-1}\label{Eq:XsDust}
\end{equation}
ionization fraction is inversely proportional to the number density. Dependences (\ref{Eq:XsRad}) and (\ref{Eq:XsDust}) can be generalized as
\begin{equation}
	x_{s} = x_0\left(\frac{n}{n_0}\right)^{-q},\label{Eq:X_n}
\end{equation}
where $q=1/2$ in the case of radiative recombinations ($x_r$), $q = 1$ in the case of recombinations on dust grains ($x_d$). 

Temperatures around the inner boundary of the accretion disk may be high enough to enable thermal ionization of metals and hydrogen. Thermal ionization fraction $x_{j}^{T}$ of the $j$-th element is calculated from Saha equation
\begin{equation}
	x\frac{x_{j}^T}{1 - x_{j}^T} = \frac{1}{n}\frac{g_j^{+}}{g_j^0}\frac{2\left(2\pi m_e kT\right)^{3/2}}{h^3} \exp\left(-\frac{\chi_j}{kT}\right),\label{Eq:ThIon}
\end{equation}
where 
\begin{equation}
	x = x_s + \sum\limits_{j}\nu_j x_j^T
\end{equation}
-- total ionization fraction, $g_{j}^0$ and $g_{j}^{+}$ -- statistical weights of neutral and ionized atoms of type $j$, $m_e$ -- electron mass, $k$ -- Boltzman constant, $h$ -- Planck constant, $\nu_j = \left(n_j^0 + n_j^{+}\right)/n_{H} = 10^{L_j - 12}$ and $\chi_{j}$-- relative abundance and ionization potential of $j$-th element, respectively, $L_{j}$ -- corresponding abundance logarithm. ``Mean'' metal with ${\chi_{\lti{Me}}=5.76\,\mathrm{eV}}$ and $L_{\lti{Me}}=5.97$ is considered for simplicity. Ionization potential and abundance logarithm of ``mean'' metal are calculated as the weighted average of corresponding parameters of Potassium, Sodium, Magnesium, Calcium and Aluminium \citep{dud87}. In addition, we take into account thermal ionization of hydrogen and helium. In the case of thermal ionization of Potassium, we get the following analytical solution of the Saha equation
\begin{eqnarray}
	x_{t} &=& 1.8\times10^{-11}\left(\frac{T}{1000\,\mbox{K}}\right)^{3/4}\left(\frac{\nu_{\lti{K}}}{10^{-7}}\right)^{0.5} \nonumber\\
	& & \times \left(\frac{n}{10^{13}\,\mbox{cm}^{-3}}\right)^{-0.5}\frac{\exp{\left(-25000 / T\right)}}{1.15\times10^{-11}}.\label{Eq:XtK}
\end{eqnarray}

\subsubsection{Magnetic field}

Radial advection and diffusion of $B_r$ and $B_{\varphi}$ are negligible in the geometrically thin disk. Condition $\mbox{div}\,{\bf B} = 0$ gives that $B_z$ does not depend on $z$ in adopted approximations. Then equations (\ref{Eq:Br}-\ref{Eq:Bz}) are transformed in the stationary case to
\begin{eqnarray}
	\frac{\partial}{\partial z}\left(V_r B_z\right) &=& - \eta\frac{\partial^2 B_r}{\partial z^2},\label{Eq:BrStat}\\
	\frac{\partial}{\partial z}\left(V_{\varphi} B_z\right) &=& - \eta\frac{\partial^2 B_{\varphi}}{\partial z^2},\label{Eq:BphiStat}\\
	\frac{\partial \left(rV_{r}B_{z}\right)}{\partial r} & = & \eta\frac{\partial}{\partial r}\left(r\frac{\partial B_{z}}{\partial r}.\right)\label{Eq:BzStat}
\end{eqnarray}
Anisotropy of the magnetic ambipolar diffusion is determined by the derivatives in the equations (\ref{Eq:BrStat}-\ref{Eq:BzStat}). Equations (\ref{Eq:BrStat}-\ref{Eq:BphiStat}) imply that $B_r$ and $B_{\varphi}$ are determined from balance between advection of $B_z$ in $r$ and $\varphi$ directions and diffusion of $B_r$ and $B_{\varphi}$ in $z$ direction. Magnetic ambipolar diffusion of $B_z$ takes place in $r$ direction, as equation (\ref{Eq:BzStat}) shows. Fossil magnetic field has equatorial symmetry so that $B_r(z=0)=0$ and $B_{\varphi}(z=0)=0$. Then, solution of (\ref{Eq:BrStat}-\ref{Eq:BphiStat}) in the thin disk approximation can be written as
\begin{eqnarray}
	B_r &=& -\frac{V_r z}{\eta}B_z,\label{Eq:BrSol}\\
	B_{\varphi} &=& -\frac{3}{2}\left(\frac{z}{r}\right)^2\frac{V_{\varphi} z}{\eta}B_z\label{Eq:BphiSol}.
\end{eqnarray}
Expressions (\ref{Eq:BrSol}-\ref{Eq:BphiSol}) evaluate radial and azimuthal components of magnetic field at height $z$ above the midplane.
 
Equations (\ref{Eq:BrSol}-\ref{Eq:BphiSol}) together with the equation (\ref{Eq:VrVphi}) imply that
\begin{equation}
	\frac{B_r}{B_{\varphi}} = \frac{2}{3}\alpha.\label{Eq:BrBphi}
\end{equation}

In previous works \citep{dud13a, dud13b} we have proposed that vertical magnetic field component in accretion disk is frozen-in. Neglecting diffusion of $B_z$ in $r$ direction, we derive from (\ref{Eq:Bz})
\begin{equation}
	rV_r B_z = \mbox{const}.\label{Eq:rVrBz}
\end{equation}
Equations (\ref{Eq:rVrBz}) and (\ref{Eq:Mass}) give
\begin{equation}
	B_z = B_{z0}\frac{\Sigma}{\Sigma_0},\label{Eq:BzFrozen}
\end{equation}
where $B_{z0}$ and $\Sigma_0$ are some typical values determined by initial conditions. Let us assume that accretion disk is formed in process of magnetostatic contraction of protostellar cloud. Then initial magnetic induction in the accretion disk may be estimated using dependence \citep[e.g.,][]{dudorov91}
\begin{equation}
	B = B_{\lti{c}}\left(\frac{\rho}{\rho_{\lti{c}}}\right)^{1/2},
\end{equation}
where $B_{\lti{c}}$ and $\rho_{\lti{c}}$ are initial magnetic induction and density of protostellar cloud. Typical values are $B_{\lti{c}} = 10^{-5}$ Gs, $n_{\lti{c}} = 10^5\,\mbox{cm}^{-3}$ (\citet{crutcher04}), so estimation of frozen-in magnetic field at the distance 1 AU is (adopting $\rho \simeq 10^{-13}\,\mbox{g}\,\mbox{cm}^{-3}$)
\begin{equation}
B_{z0} = B_{\lti{c}}\left(\rho(1\,\mbox{AU})/\rho_{\lti{c}}\right)^{1/2} = 0.16\,\mathrm{Gs}. 
\end{equation}

Solution (\ref{Eq:BzFrozen}) is valid only in the regions of high ionization fraction, where magnetic diffusion is inefficient. MAD is efficient if MAD time scale $\tau_{\lti{ad}}$ is of the order of magnetic field generation time scale $\tau_{z}$ \citep{nakano72}. Vertical magnetic field component is amplified due to accretion with the velocity $V_r$. Magnetic ambipolar diffusion of the vertical magnetic field component takes place in the $r$ direction. Thus, we define $\tau_{z} = r/V_r$ and $\tau_{\lti{ad}} = r / (V_{\lti{ad}})_r$, so that
\begin{equation}
	\left(\tau_{\lti{ad}}\right)_r \simeq \frac{4\pi x\rho^2\eta_{\lti{in}}r^2}{B^2}. \label{Eq:Tmad}
\end{equation}
We get from equality between $\tau_z$ and $\tau_{\lti{ad}}$ (or, equivalently, velocities $V_r$ and $(V_{\lti{ad}})_r$)
\begin{equation}
	B_z = \left(4\pi\eta_{\lti{in}} x\rho^2 r\right)^{1/2}. \label{Eq:BzMAD}
\end{equation}
We derive power-law dependence of $B_z$ on density from (\ref{Eq:BzMAD}) using (\ref{Eq:VrVphi}) and (\ref{Eq:X_n}),
\begin{equation}
	B_z = C_B \rho^{1-q/2}\alpha^{1/2}M^{1/4}\left(\frac{H}{r}\right)r^{1/4}, \label{Eq:BzMADq}
\end{equation}
where
\begin{equation}
	C_B = \left(4\pi\eta_{in} x_0 (n_0 m_p\mu)^{q}\sqrt{G}\right)^{1/2}.
\end{equation}
For example, formula (\ref{Eq:BzMADq}) gives in the case of recombinations on dust grains at $\xi=10^{-17}\,\mathrm{s}^{-1}$, $\alpha_{g}=\alpha_{g0}$
\begin{eqnarray}
	B_z &=& 0.013 \, \left(\frac{\rho}{10^{-13}\,\mbox{g}\,\mbox{cm}^{-3}}\right)^{1/2} \left(\frac{r}{1\,\mbox{AU}}\right)^{1/4}\nonumber\\
	& & \times \left(\frac{\alpha}{0.01}\right)^{1/2} \left(\frac{M}{M_{\odot}}\right)^{1/4} \left(\frac{H/r}{0.05}\right)\,\mbox{Gs},\label{Eq:BzAnalytic}
\end{eqnarray}
i.e. MAD leads to nearly order of magnitude reduction of $B_z$ at 1 AU comparing to the frozen-in magnetic field. 

Ohmic dissipation also operates in the regions of low ionization fraction. Time scale of OD in radial direction $\tau_{OD}=r^2/\nu_m$ is
\begin{eqnarray}
	\tau_{OD} &=& 16\,\left(\frac{\xi}{10^{-17}\,\mbox{s}^{-1}}\right)\left(\frac{n}{10^{13}\,\mbox{cm}^{-3}}\right)^{-1} \nonumber\\
	& & \times \left(\frac{T}{400\,\mbox{K}}\right)^{-1/2}\left(\frac{r}{1\,\mbox{AU}}\right)^{2}\,\mbox{yr}, \label{Eq:tauOD}
\end{eqnarray}
in the case of recombinations on dust grains at $\xi=10^{-17}\,\mathrm{s}^{-1}$, $\alpha_{g}=\alpha_{g0}$. According to this estimation, OD can be efficient in accretion disks. We believe that efficient OD of $B_z$ field prevents amplification of magnetic field in regions where $\tau_{OD}$ is lower than accretion disk life time $t_{disk}$, so estimation (\ref{Eq:BzMAD}) remains valid.

\section{Analytical solution}
\label{Sec:AnalytResults}
Equations system (\ref{Eq:Kepler}, \ref{Eq:H}, \ref{Eq:Mass}-\ref{Eq:Energy}, \ref{Eq:Teff}, \ref{Eq:TirrAppr}, \ref{Eq:Ion}, \ref{Eq:ThIon}, \ref{Eq:BrSol}, \ref{Eq:BphiSol}, \ref{Eq:BzFrozen}, \ref{Eq:BzMAD}) describes the dynamics of stationary accretion disk with fossil large-scale magnetic field in the kinematic approximation. Parameters of the model are: $\alpha$, $\dot{M}$, physical characteristics of the star: mass $M$, radius $R_{s}$, surface magnetic field $B_s$, luminosity $L_{\star}$, opacity parameters: $a$, $b$ and $\kappa_0$, dust-to-gas mass ratio $Y_{\lti{g}}$, dust size and ionization rates.

This system of model equations is closed and it has analytical solution if we use power-law dependence of ionization fraction on density (\ref{Eq:X_n}) instead of solution of equations (\ref{Eq:Ion}, \ref{Eq:ThIon}).

We use following parameters as fiducial: ${\alpha = 0.01}$, $M=1\,M_{\odot}$, ${\dot{ M}=10^{-8}\,M_{\odot}/\mbox{yr}}$, ${\xi=10^{-17}\,\mathrm{s}^{-1}}$, $\alpha_{g}=\alpha_{g0}$ and ${L_{\lti{s}}=1\,L_{\odot}}$. At the low temperatures ($T<1000$ K) opacity parameters are (see table \ref{Tab:1}): $a=0,\,b=1,\, \kappa_0 = 3\times 10^{-3} \, \mbox{cm}^2/\mbox{g}$. With the help of non-dimensional variables
\begin{eqnarray}
	& & \alpha_{0.01}=\frac{\alpha}{0.01},\, \dot{m} = \frac{\dot{M}}{10^{-8}\,M_{\odot}/\mbox{yr}},\, m=\frac{M}{M_{\odot}} \nonumber\\
	& & r_{\lti{AU}} = \frac{r}{1\,\mbox{AU}}, \quad h = \frac{H}{1\,\mbox{AU}},\, \xi_{-17}=\frac{\xi}{10^{-17}\,\mathrm{s}^{-1}}
\end{eqnarray}
we derive analytical solution describing accretion disk radial structure for $f=1$,
{
\allowdisplaybreaks
\begin{eqnarray}
  T_{\lti{eff}} &=& 150\, \dot{m}^{1/4} m^{1/4} r_{\lti{AU}}^{-3/4}\,\mbox{K},\label{Eq:TeAppr}\\
  T_{\lti{c}} &=& 240 \, \alpha_{0.01}^{-1/4} \dot m^{1/2} m^{3/8} r_{\lti{AU}}^{-9/8}\, \mbox{K},\label{Eq:TcAppr}\\
  \Sigma &=& 230 \, \alpha_{0.01}^{-3/4} \dot{m}^{1/2} m^{1/8} r_{\lti{AU}}^{-3/8}\, \mbox{g} \,\mbox{cm}^{-2},\label{Eq:SAppr}\\
  \rho &=& 2.5\times 10^{-10} \, \alpha_{0.01}^{-5/8} \dot{m}^{1/4} m^{7/16} \nonumber\\
  & & \times r_{\lti{AU}}^{-21/16}\, \mbox{g}\, \mbox{cm}^{-3},\label{Eq:RhoAppr}\\
  h &=& 0.03\, \alpha_{0.01}^{-1/8} \dot{m}^{1/4} m^{-5/16} r_{\lti{AU}}^{15/16}\, \mbox{AU},\label{Eq:HAppr}\\
  V_r &=& - 30 \, \alpha_{0.01}^{3/4} \dot{m}^{1/2} m^{-1/8} r_{\lti{AU}}^{-5/8}\, \mbox{cm}\, \mbox{s}^{-1},\label{Eq:VrAppr}\\
	x_{g} &=& 3.4\times 10^{-15}\,\xi_{-17}\alpha_{0.01}^{5/8} \dot{m}^{-7/16} m^{-1/4}\nonumber\\
	& &\times r_{\lti{AU}}^{21/16},\label{Eq:XdAppr}\\
	x_{r} &=& 2.0\times 10^{-10}\,\xi_{-17}^{1/2}\alpha_{0.01}^{1/4} m^{-1/8} r_{\lti{AU}}^{3/8},\label{Eq:XrAppr}\\
	B_z &=& 0.29\,\alpha_{0.01}^{-3/4} \dot{m}^{1/2} m^{1/8} r_{\lti{AU}}^{-3/8}\, \mbox{Gs},\label{Eq:BzApprVisc}\\
	B_z^{\lti{mad}} &=& 0.024\,\xi_{-17}^{1/2}\alpha_{0.01}^{1/16} \dot{m}^{3/8} m^{5/32} r_{\lti{AU}}^{-15/32}\, \mbox{Gs},\label{Eq:BzMadApprVisc}\\
	B_r^{\lti{od}} &=& 1.5\times 10^{-7}\,\xi_{-17}^{3/2}\alpha_{0.01}^{23/16} \dot{m}^{5/8} m^{-29/32} \nonumber\\
	& & \times r_{\lti{AU}}^{55/32}\, \mbox{Gs},\label{Eq:BrODApprVisc}\\
	B_{\varphi}^{\lti{od}} &=& 2.2\times 10^{-5}\,\xi_{-17}^{3/2}\alpha_{0.01}^{7/16} \dot{m}^{5/8} m^{-29/32} \nonumber\\
	& & \times r_{\lti{AU}}^{55/32}\, \mbox{Gs},\label{Eq:BphiODApprVisc}\\
	B_r^{\lti{mad}} &=& 7.4\times 10^{-4}\,\xi_{-17}^{1/2}\alpha_{0.01}^{-1/16} \dot{m}^{5/8} m^{-5/32} \nonumber\\
	& & \times r_{\lti{AU}}^{-17/32}\, \mbox{Gs},\label{Eq:BrMADApprVisc}\\
	B_{\varphi}^{\lti{mad}} &=& 0.11\,\xi_{-17}^{1/2}\alpha_{0.01}^{-17/16} \dot{m}^{5/8} m^{-5/32}\nonumber\\
	& & \times r_{\lti{AU}}^{-17/32}\, \mbox{Gs},\label{Eq:BphiMADApprVisc}
\end{eqnarray}}
Accretion disk density is determined as $\rho=\Sigma/\left(2H\right)$ here. Ionization fraction profiles (\ref{Eq:XdAppr}) and (\ref{Eq:XrAppr}) correspond to the case of recombinations on the dust grains and radiative recombinations, respectively. Dependence (\ref{Eq:BzApprVisc}) is obtained using equation (\ref{Eq:BzFrozen}) for the frozen-in field. Radial profiles of the magnetic field components (\ref{Eq:BzMadApprVisc}-\ref{Eq:BphiMADApprVisc}) are obtained for the case of recombinations on the dust grains, $x=x_g$. Dependence (\ref{Eq:BzMadApprVisc}) is derived according to equation (\ref{Eq:BzMAD}). Profiles of the radial and azimuthal magnetic field components (\ref{Eq:BrODApprVisc}-\ref{Eq:BphiMADApprVisc}) are obtained from equations (\ref{Eq:BrSol}, \ref{Eq:BphiSol}, \ref{Eq:BzMAD}) using OD coefficient ($B_r^{\lti{od}}$ and $B_{\varphi}^{\lti{od}}$) and MAD coefficient ($B_r^{\lti{mad}}$ and $B_{\varphi}^{\lti{mad}}$). Comparison of these dependences allows us to find out the relative importance of certain magnetic diffusion type.

We derive solution for irradiated disk from equations (\ref{Eq:Kepler}, \ref{Eq:H}, \ref{Eq:Mass}, \ref{Eq:AngMom}, \ref{Eq:X_n}, \ref{Eq:BrSol}, \ref{Eq:BphiSol}, \ref{Eq:BzFrozen}, \ref{Eq:BzMAD}) using temperature profile (\ref{Eq:TirrAppr}) instead of solution of equation (\ref{Eq:Energy}),
{
\allowdisplaybreaks
\begin{eqnarray}
	T &=& 280\,l_{s}^{1/4} r_{\lti{AU}}^{-1/2}\,\mbox{K},\label{Eq:TirrSol}\\
	\Sigma &=& 400 \, \alpha_{0.01}^{-1} \dot{m} m^{1/2} r_{\lti{AU}}^{-1} l_{\lti{s}}^{-1/4}\, \mbox{g}\, \mbox{cm}^{-2}\label{Eq:Sirr} \\ 
	\rho &=& 1.96\times 10^{-10} \, \alpha_{0.01}^{-1} \dot{m} m r_{\lti{AU}}^{-9/4} l_{\lti{s}}^{-3/4}\, \mbox{g}\, \mbox{cm}^{-3} \label{Eq:RhoIrrAppr}\\ 
	h &=& 0.03 \, m^{-1/2} r_{\lti{AU}}^{5/4} l_{\lti{s}}^{1/8}\, \mbox{AU}, \label{Eq:HApprIrr}\\
	V_r &=& -34 \, \alpha_{0.01} m^{-1/2} l_{\lti{s}}^{1/4}\, \mbox{cm}\, \mbox{s}^{-1},\\
	x_{d} &=& 4.4\times 10^{-15}\,\xi_{-17}\alpha_{0.01} \dot{m}^{-1} m^{-1} l_{\lti{s}}^{3/8}r_{\lti{AU}}^{9/4},\label{Eq:XdIrrAppr}\\
	x_{r} &=& 1.5\times 10^{-10}\,\xi_{-17}^{1/2}\alpha_{0.01}^{1/2}\dot{m}^{-1/2} m^{-1/2} \nonumber\\
	& & \times l_{\lti{s}}^{1/4}r_{\lti{AU}},\label{Eq:XrIrrAppr}\\
	B_z &=& 0.29\,\alpha_{0.01}^{-1} \dot{m} m^{1/2} r_{\lti{AU}}^{-1}\, \mbox{Gs},\label{Eq:BzApprIrr}\\
	B_z^{\lti{mad}} &=& 0.023\,\xi_{-17}^{1/2} \dot{m}^{1/2} m^{1/4} \nonumber\\
	& & \times l_{\lti{s}}^{1/16} r_{\lti{AU}}^{-5/8}\, \mbox{Gs},\label{Eq:BzMADApprIrr}\\
	B_r^{\lti{od}} &=& 2.2\times 10^{-7}\,\xi_{-17}^{3/2}\alpha_{0.01}^{2} \dot{m}^{-1/2} m^{-7/4} \nonumber\\
	& & \times l_{\lti{s}}^{9/16}r_{\lti{AU}}^{25/8}\, \mbox{Gs},\label{Eq:BrODApprIrr}\\
	B_{\varphi}^{\lti{od}} &=& 3.2\times 10^{-5}\,\xi_{-17}^{3/2}\alpha_{0.01} \dot{m}^{-1/2} m^{-7/4} \nonumber\\
	& & \times l_{\lti{s}}^{9/16} r_{\lti{AU}}^{25/8}\, \mbox{Gs},\label{Eq:BphiODApprIrr}\\
	B_r^{\lti{mad}} &=& 0.006\,\xi_{-17}^{1/2} \dot{m}^{1/2} m^{-1/4} \nonumber\\
	& & \times l_{\lti{s}}^{1/16} r_{\lti{AU}}^{-3/8}\, \mbox{Gs},\label{Eq:BrMADApprIrr}\\
	B_{\varphi}^{\lti{mad}} &=& 0.11\,\xi_{-17}^{1/2}\alpha_{0.01}^{-1} \dot{m}^{1/2} m^{-1/4} \nonumber\\
	& & \times l_{\lti{s}}^{1/16} r_{\lti{AU}}^{-3/8}\, \mbox{Gs},\label{Eq:BphiMADApprIrr}
\end{eqnarray}}
where $l_{\lti{s}} = L_{\star}/L_{\odot}$.

Power-law solution (\ref{Eq:TeAppr}-\ref{Eq:VrAppr}) is analogous to those derived by \cite{ss73} for viscous accretion disks around black holes. Accretion disks of young stars are so cold that dust grains make main contribution to the opacity.
Temperature in accretion disk of young stars is larger than dust evaporation temperature ${\sim 2\times 10^3\,\mathrm{K}}$ only at ${r \lesssim 0.15\,\mathrm{AU}}$ according to (\ref{Eq:TcAppr}) for fiducial parameters. Opacities due to molecules dissociation and atoms ionization dominate here (rows 7-9 in the table \ref{Tab:1}). Solution similar to (\ref{Eq:TeAppr}-\ref{Eq:VrAppr}) can be derived for gas opacity dominated region, but we do not present it here because the corresponding region is very small. 

Equations of the model are non-linear if we take into account thermal ionization, and analytical solution for the ionization fraction and magnetic field components cannot be obtained. Ionization fraction is large near the accretion disk inner boundary where thermal ionization of metals operates: ${x>10^{-5}}$ at ${r< r_{x_{T}}}$ according to (\ref{Eq:XtK}), where ${r_{x_{T}} \equiv r\left(10^3\,\mbox{K}\right) \simeq 0.3\,\mathrm{AU}}$ as it follows from (\ref{Eq:TcAppr}). Magnetic field diffusion is inefficient in this region.

Accretion disks are cold (${T<2\times 10^3\,\mathrm{K}}$) at the distances ${r>0.15\,\mathrm{AU}}$, so radiation transfer is determined by dust opacities (rows 1-6 in the table \ref{Tab:1}). Comparison of (\ref{Eq:TcAppr}) and (\ref{Eq:TirrSol}) shows that irradiation temperature of the accretion disk is higher than temperature due to viscous heating at the distances further than
\begin{equation}
	r_{\lti{v}} = 0.8\,\alpha_{0.01}^{-2/5}\dot{m}^{4/5}m^{3/5}l_{s}^{-2/5}\,\mbox{AU}.
\end{equation}

Viscous heating dominates at $r<r_{\lti{v}}$. Solution (\ref{Eq:TeAppr}-\ref{Eq:BphiMADApprVisc}) describes this region. Region where stellar radiation is the dominant heating mechanism, $r>r_{\lti{v}}$, is characterized by solution (\ref{Eq:TirrSol}-\ref{Eq:BphiMADApprIrr}).

Slope of the surface density radial profile at ${r<r_{\lti{v}}}$ is quiet flat, ${\Sigma \propto r^{-3/8}}$, in comparison with very steep profile in MMSN, ${\Sigma \propto r^{-3/2}}$. Surface density in our solution, ${\Sigma(1\,\mbox{AU})=230\,\mbox{g}\,\mbox{cm}^{-2}}$, is less than standard value ${1700\,\mbox{g}\,\mbox{cm}^{-2}}$ in MMSN. Surface density slope is more steep, ${\Sigma \propto r^{-1}}$, at ${r>r_{\lti{v}}}$. Scale height radial profiles (\ref{Eq:HAppr}) and (\ref{Eq:HApprIrr}) show that  ${H/r \approx 0.03 \approx \mbox{const}}$. This result confirms our assumption that the accretion disk is geometrically thin, ${H/r \ll 1}$.

We determine accretion disk outer boundary $r_{\lti{out}}$ as the contact boundary. Density in molecular cloud cores ${n\approx 10^9\,\mbox{cm}^{-3}}$ and temperatures ${T\simeq 20\,\mathrm{K}}$ \citep{andre93}. It follows from (\ref{Eq:TirrSol}) and (\ref{Eq:RhoIrrAppr}) that ${r_{\lti{out}}\sim 140\,\mathrm{AU}}$ for the accretion disk of solar mass star.

Equation (\ref{Eq:SAppr}) shows that surface density is more than attenuation length of the cosmic rays ${\sim 100\,\mbox{g}/\mbox{cm}^2}$ at distances less than several AU. Therefore, ionization fraction is very small ${x_g(1\,\mbox{AU})\simeq 10^{-14}}$ according to (\ref{Eq:XdAppr}). Density decreases with distance (\ref{Eq:RhoAppr}, \ref{Eq:RhoIrrAppr}) that leads to increase of the ionization fraction. Ionization fraction profile slope is more steep in the case of recombinations on the dust grains (\ref{Eq:XdAppr}, \ref{Eq:XdIrrAppr}) than in the radiative recombinations case (\ref{Eq:XrAppr}, \ref{Eq:XrIrrAppr}). Equations (\ref{Eq:XdAppr}) and (\ref{Eq:XrAppr}) show that ionization fraction at ${r=1\,\mathrm{AU}}$ is 5 orders of magnitude smaller in the case of recombinations on the dust grains than in the pure radiative recombinations case. Ionization fraction reaches $10^{-10}$ near the outer boundary in the case of recombinations on the dust grains (\ref{Eq:XdIrrAppr}), while ${x(r_{\lti{out}})\simeq 2\times 10^{-8}}$ in the case of radiative recombinations (\ref{Eq:XrIrrAppr}).

Frozen-in vertical component of the magnetic field depends on $r$ as ${B_z = 0.29\,r_{\lti{AU}}^{-3/8}\,\mathrm{Gs}}$ (\ref{Eq:BzApprVisc}) at ${r<r_{\lti{v}}}$. Ionization fraction is very low	, ${x<10^{-14}}$, at ${r_{x_T} < r < r_{\lti{v}}}$ according to (\ref{Eq:XdAppr}). Effective MAD leads to reduction of $B_z$ by order of magnitude comparing to frozen-in field at ${r=1\,\mathrm{AU}}$, so that ${B_z = 0.024\,r_{\lti{AU}}^{-15/32}\,\mathrm{Gs}}$ (\ref{Eq:BzMadApprVisc}). Comparison of dependences (\ref{Eq:BrODApprVisc}), (\ref{Eq:BphiODApprVisc}) and (\ref{Eq:BrMADApprVisc}), (\ref{Eq:BphiMADApprVisc}) show that OD is the main magnetic diffusion mechanism at ${r_{x_T} < r < r_{\lti{v}}}$. Ohmic diffusion is more efficient than MAD at ${r < r_{\lti{OD}} \simeq 14\,\mathrm{AU}}$ according to dependences (\ref{Eq:BphiODApprIrr}) and (\ref{Eq:BphiMADApprIrr}). Magnetic field is quasi-poloidal in this region, ${B_z \gg B_{\varphi} \gg B_r}$ at ${r_{x_T} < r < r_{\lti{OD}}}$.

Magnetic ambipolar diffusion is the main magnetic diffusion mechanism at distances ${r > r_{\lti{OD}}}$, $B_r$ and $B_{\varphi}$ decrease with distance here: ${B_r = 0.011\,r_{\lti{AU}}^{-3/8}\,\mathrm{Gs}}$,  ${B_{\varphi} = 0.2\,r_{\lti{AU}}^{-3/8}\,\mathrm{Gs}}$. Comparison of the profiles (\ref{Eq:BzApprIrr}) and (\ref{Eq:BzMADApprIrr}) shows that MAD becomes inefficient near $r_{\lti{out}}$ where ${x \gtrsim 10^{-10}}$ and ${B_{r}(r_{\lti{out}}) \simeq 1\,\mathrm{mGs}}$, $B_{\varphi}(r_{\lti{out}}) \simeq 17\,\mathrm{mGs}$ and ${B_{z}(r_{\lti{out}}) \simeq 2\,\mathrm{mGs}}$. Hence, magnetic field has quasi-azimuthal geometry (${B_{\varphi} > B_z > B_r}$) in the outer region of the accretion disk.

\section{Numerical solution}
\label{Sec:NumResults}
System of equations (\ref{Eq:Kepler}, \ref{Eq:H}, \ref{Eq:EnergyStat}-\ref{Eq:Mass}, \ref{Eq:Ion}, \ref{Eq:ThIon}, \ref{Eq:BrSol}, \ref{Eq:BphiSol}, \ref{Eq:BzFrozen}, \ref{Eq:BzMAD}) is essentially non-linear when OD and MAD, shock and thermal ionization are taken into account simultaneously. We solve these equations numerically in the distance range between $r_{\lti{in}}$ and $r_{\lti{out}}$ on the log-scaled space grid. Accretion disk inner boundary $r_{\lti{in}}$ is determined by the radius of the magnetosphere that is calculated from the balance between viscous and Maxwell stresses $\rho V_r V_{\varphi} = B_z B_{\varphi} / (4\pi)$,
\begin{eqnarray}
	\frac{r_{\lti{in}}}{2R_{\odot}} &=& 2.9 \, \left(\frac{B_{\lti{s}}}{2\,\mbox{kGs}}\right)^{4/7} \, \left(\frac{R_{\lti{s}}}{2\,R_{\odot}}\right)^{12/7} \nonumber \\
	& & \times\left(\frac{\dot M}{10^{-7}\,M_{\odot}/\mbox{yr}}\right)^{-2/7} \, \left(\frac{M}{M_{\odot}}\right)^{-1/7},\label{Eq:Rin}
\end{eqnarray}
where it is adopted that $H/r =0.1$ and $B_{s} \sim B_{\varphi}\sim B_z$ at $r_{\lti{in}}$.

At first step of calculation, we determine temperature $T$, surface density $\Sigma$, scale height $H$ and radial velocity $V_r$ by solving equations (\ref{Eq:Kepler}, \ref{Eq:H}, \ref{Eq:Mass}-\ref{Eq:Energy}) with opacity parameters from the Table \ref{Tab:1}. We use a simple approach in order to take into account stellar irradiation. If midplane temperature calculated from (\ref{Eq:Energy}) is lower than the irradiation temperature (\ref{Eq:TirrSol}) then $T$ is replaced by $T_{\lti{irr}}$. At the next step, we substitute temperature and density into the ionization subsystem (\ref{Eq:Ion}, \ref{Eq:ThIon}), that is solved using the iterative Newton method. 

After calculation of ionization fraction $x$ and conductivity $\sigma$, we determine the vertical magnetic field component from the equations (\ref{Eq:BzFrozen}, \ref{Eq:BzMAD}). Finally, radial and azimuthal magnetic field components are calculated from (\ref{Eq:BrSol}, \ref{Eq:BphiSol}). We use coefficient of MAD in the form $\eta_{\lti{ad}}=B_z^2/(4\pi x\rho^2\eta_{\lti{in}})$, so equations (\ref{Eq:BrSol}, \ref{Eq:BphiSol}) are linear. As it was found in the previous section, OD and MAD are inefficient in the innermost thermally ionized region of the accretion disk ($r\lesssim 0.3$ AU). Differential rotation leads to generation of the strong azimuthal magnetic field component in this region. We propose that buoyancy limits $B_{\varphi}$. According to \cite{parker_book}, strength of the azimuthal component of the magnetic field is determined from equality of rise time of azimuthal magnetic flux tubes and time of magnetic field generation. We assume that terminal velocity of rise is equal to the Alfven velocity $V_{a\varphi}=B_{\varphi}/\sqrt{4\pi\rho}$ and corresponding time is $t_b=H/V_{a\varphi}$. Time of $B_{\varphi}$ generation up to value of $B_z$ follows from (\ref{Eq:Bphi})
\begin{equation}
t_{\varphi}=\frac{2}{3}\Omega_{\lti{k}}^{-1}\left(\frac{H}{r}\right)^{-1},\label{Eq:Bphi_time}
\end{equation}
i.e. it is several times larger than rotation period $2\pi/\Omega_{\lti{k}}$. Equality of $t_b$ and $t_{\varphi}$ gives
\begin{equation}
	B_{\varphi}= 1.5(H/r)V_{\lti{s}}\sqrt{4\pi\rho}.\label{Eq:BphiMax}
\end{equation}
We assume that azimuthal component of the magnetic field is determined from (\ref{Eq:BphiMax}) in the regions where OD and MAD are ineffective.

Following parameters are used in our calculations: $\alpha=0.01$, $B_s = 2$ kGs, $R_s = 2\,R_{\odot}$ \citep{yang11}. We carry out calculations for stellar masses $M=0.5-2\,M_{\odot}$ and use observational correlations of stellar luminosity and accretion rate with stellar mass. Dimensionless accretion rate, $\dot{m} = m^2$ \citep{fang09}.  Accretion disk mass is calculated as
\begin{equation}
	M_{\lti{disk}} = 2\pi \int\limits_{r_{\lti{in}}}^{r_{\lti{out}}}r\Sigma_{\lti{tot}}dr,\label{Eq:Mdisk}
\end{equation}
where $\Sigma_{\lti{tot}}=\int\limits_{-\infty}^{\infty}\rho dz=\Sigma / \mbox{erf}(1/\sqrt{2})$ -- total surface density. Definitions of accretion disk mass (\ref{Eq:Mdisk}), inner boundary (\ref{Eq:Rin}) and outer boundary (contact boundary) together with equations (\ref{Eq:TirrSol}-\ref{Eq:RhoIrrAppr}) yield that $M_{\lti{disk}}=0.027\,M_{\odot}$, $r_{\lti{in}}=0.052$ AU and $r_{\lti{out}}=140$ AU at the fiducial parameters.

\subsection{Accretion disk structure}
\label{Sect:Disk}

\begin{figure*}[!htb]
\centering
\includegraphics[width=0.98\textwidth, trim=0cm 0cm 0cm 0cm]{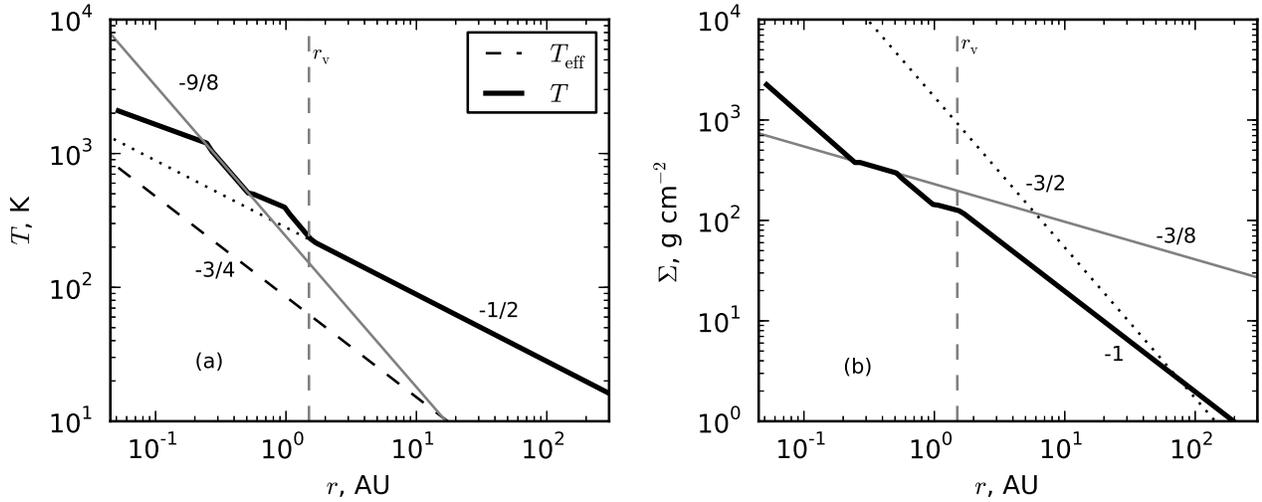}
\caption{{\it Panel a}: Radial profile of the temperature in the accretion disk of solar mass star. Solid black line: numerical solution, dashed line: effective temperature, gray solid line: analytical solution (\ref{Eq:TcAppr}) describing the region of the dominant viscous heating, dots: analytical solution (\ref{Eq:TirrSol}) describing the region where stellar irradiation is the main heating source. {\it Panel b}: Radial profile of surface density. Solid black line: numerical solution, dots: the profile of the MMSN model, gray solid line: analytical solution (\ref{Eq:SAppr}) for dominant viscous heating. Numbers near the curves denote slopes. Vertical dashed gray line $r_{\lti{v}}$ separates region of dominant viscous and irradiation heating.}
\label{Fig:1}
\end{figure*}

In figure \ref{Fig:1}, we plot accretion disk temperature (panel a) and surface density (panel b) calculated for the fiducial parameters. Figure \ref{Fig:1}(a) shows that accretion disk temperature is determined by viscous heating at distances less than $r_{\lti{v}} \sim 2$ AU, and by stellar irradiation at larger distances. Variation of slope of temperature radial profile at $r < r_{\lti{v}}$ is due to dust evaporation and corresponding variation of opacity. Midplane temperature $T$ is larger than the effective temperature $T_{\lti{eff}}$ at distances less than $\sim 20$ AU, because accretion disk is optically thick to its own radiation in this region.  Typical slope of $\Sigma(r)$ is $-3/8$ at $r<r_{\lti{v}}$ according to analytical solution (\ref{Eq:SAppr}). Surface density decreases with distance more rapidly, $\Sigma \propto r^{-1}$ in region of dominant heating by stellar radiation, $r > r_{\lti{v}}$, in accordance with the analytical solution (\ref{Eq:Sirr}). Typical surface density slopes agree with observational ones \citep{williams11}. We note that the surface density profile introduced in MMSN model is much steeper, $\Sigma \propto r^{-3/2}$.

\subsection{Ionization fraction}
\label{Sect:Ion}

We consider shock ionization by cosmic rays, X-Rays and radionuclides. Cosmic rays ionization rate \citep{spitzer68}
\begin{equation}
	\xi_{\lti{CR}}(r,z) = \xi_0\exp\left(-\frac{\Sigma(r,z)}{R_{\lti{CR}}}\right),
\end{equation}
where
\begin{equation}
	\Sigma(r,z) = \int\limits_{z}^{\infty}\rho dz \label{Eq:SigmaZ}
\end{equation}
-- surface density for given latitude $z$, $R_{\lti{CR}}=100\,\mbox{g}\,\mbox{cm}^{-2}$ -- cosmic rays attenuation length. We use two values of cosmic rays ionization rate coefficient, $\xi_0=10^{-17}\,\mbox{s}^{-1}$ \citep{spitzer68} and $\xi_0=10^{-16}\,\mbox{s}^{-1}$ \citep[e.g.,][]{sorochenko10}.
Ionization rate by stellar X-rays is calculated using approximation from \citet{bai09},
\begin{eqnarray}
	\xi_{\lti{XR}} &=& \frac{L_{\lti{XR}}}{10^{29}\mbox{erg}\mbox{s}^{-1}}r_{\lti{AU}}^{-2.2}\left[\xi_1\exp\left(-\left(\frac{N}{N_1}\right)^{a_1}\right) \right. \nonumber\\
	& & + \left. \xi_2\exp\left(-\left(\frac{N}{N_2}\right)^{b_1}\right)\right].
\end{eqnarray}
where $L_{XR}$ is the X-ray luminosity of the star, column density $N(r,\, z)=\Sigma(r,\,z)/m_{n}$. Adopted approximation corresponds to photons energy $kT_{\lti{XR}}=3$ keV and height of the X-ray source above the midplane $R_{\lti{XR}} = 10\,R_{\odot}$, so that $\xi_1 = 6\times 10^{-12}\,\mbox{s}^{-1}$, $\xi_2 = 10^{-15}\,\mbox{s}^{-1}$, $N_1 = 1.5\times 10^{21}\,\mbox{cm}^{-2}$, $N_2 = 7\times 10^{23}\,\mbox{cm}^{-2}$, $a_1=0.4$, $b_1=0.65$. Surface density $\Sigma(r,z)$ is determined by integration of hydrostatic density profile (\ref{Eq:HydroStat}) from $z$ to $z_0 \gg z$. We vary X-rays luminosities in the range $L_{XR}=\left(10^{29}-10^{32}\right)\,\mbox{erg}\,\mbox{s}^{-1}$ according to observations \citep{casanova95}. We use following ionization parameters as a fiducial: $a_d=0.1\,\mu$m, $\xi_0=10^{-17}\,\mbox{s}^{-1}$, $L_{XR}=10^{30}\,\mbox{erg}\,\mbox{s}^{-1}$. Ionization by $^{40} K$ with the rate $\xi_{RE}=6.9\times 10^{-23}\,\mbox{s}^{-1}$ \citep{sano00} is also taken into account.

\begin{figure*}[t]
\centering
\includegraphics[width=0.98\textwidth, trim=0cm 0cm 0cm 0cm]{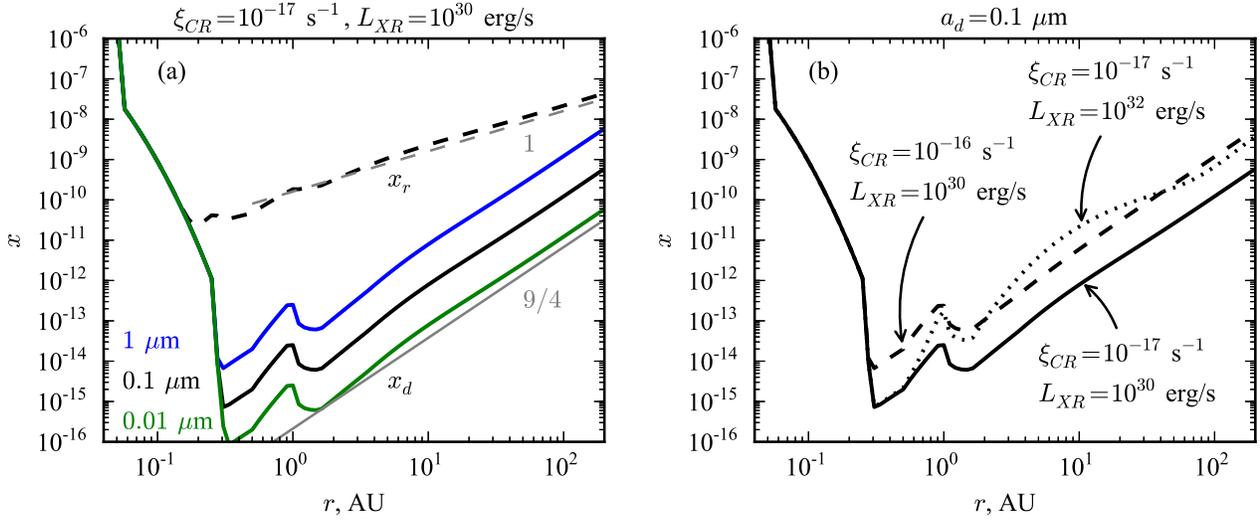}
\caption{Radial profiles of the ionization fraction in the midplane of the accretion disk of solar mass star. {\it Left}: dashed black line depicts ionization fraction in the case of radiative recombinations ($x_r$), solid lines -- recombinations on the dust grains ($x_d$, green line: $a_d=0.01\,\mu$m, black line: $a_d=0.1\,\mu$m, blue line: $a_d=1\,\mu$m). Labelled gray lines depict typical slopes of the analytical dependences (\ref{Eq:XdIrrAppr}) and (\ref{Eq:XrIrrAppr}). {\it Right}: Radial profiles of midplane ionization fraction for different ionization rates.}
\label{Fig:2}
\end{figure*}

Radial profiles of the ionization fraction in the midplane of the accretion disk of solar mass star are plotted in the panel (a) of the figure \ref{Fig:2}. Figure \ref{Fig:2}(a) shows that the ionization fraction radial profile is non-monotonous. Ionization fraction is minimal, $x_{\lti{min}} \approx 10^{-15}$ at $r\sim 0.3-0.4$ AU, in the case of recombinations on dust grains with $a_d=0.1\,\mu$m. Ionization fraction increases at larger distances, where ionization by cosmic rays and X-rays is more effective. Thermal ionization takes place at smaller distances, $r<0.3$ AU, and it leads to fast increase of $x$ approaching to the star. 

Growth of dust grain size in 10 times leads to order of magnitude growth of the ionization fraction. This is because dust grains cross-section is proportional to $a_d^2$ and dust-to-gas density ratio $X_{\lti{d}}$ is proportional to $a_d^{-3}$ so that $\alpha_g \propto a_d^{-1}$ and $x_d \propto a_d$ according to (\ref{Eq:Alpha_g}, \ref{Eq:XsDust}). Ionization fraction is larger by several orders of magnitude in the case of radiative recombinations, so that $x_{\lti{min}} \approx 10^{-11}$. Small peaks on the $x(r)$ dependence at $r \simeq 1$ AU are related to evaporation of the ice dust grains.

Midplane ionization fraction radial profiles for different ionization rates are plotted in the panel (b) of the figure \ref{Fig:2}.
Comparison of $x(r)$ dependences between each other shows that growth of cosmic rays ionization rate by order of magnitude leads to growth of midplane ionization fraction by order of magnitude (see expression (\ref{Eq:XsDust})). Midplane ionization fraction is less sensitive to X-ray luminosity. Growth of $L_{XR}$ in 100 times leads to nearly order of magnitude increase of midplane ionization fraction at the distances $r>1$ AU.

\subsection{Fossil magnetic field of accretion disk}

In this section, we calculate radial profiles of the fossil magnetic field components and analyse magnetic field geometry. Absolute values of magnetic field components are plotted in the figures.

\subsubsection{Vertical magnetic field}

\begin{figure*}[t]
\includegraphics[width=0.98\textwidth, trim=0cm 0cm 0cm 0cm]{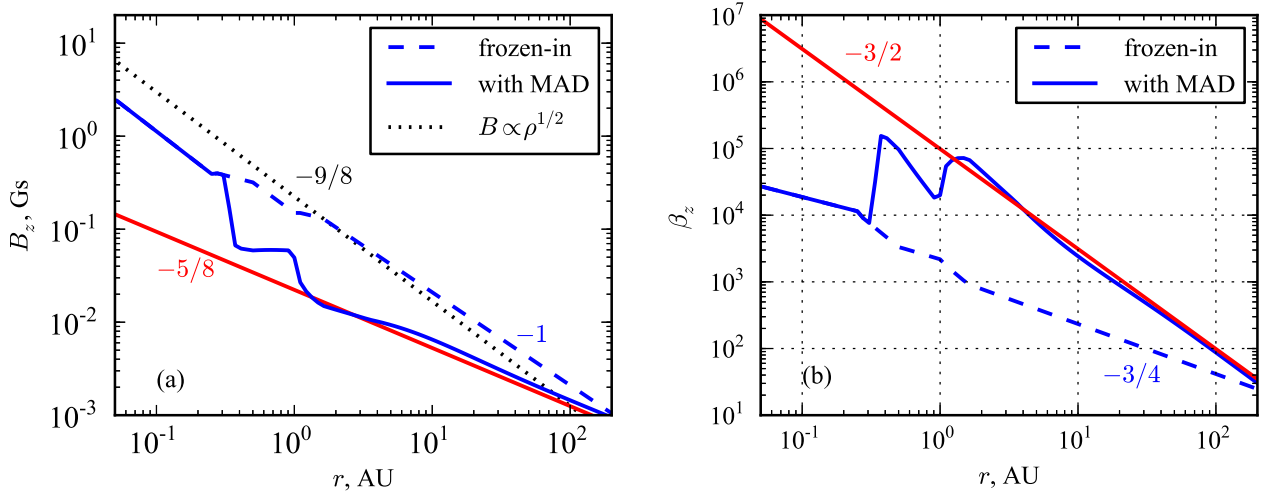}
\caption{Radial profiles of the vertical magnetic field component ({\it left}) and corresponding radial profiles of midplane plasma parameter ({\it right}) in the accretion disk of solar mass star. Blue dashed lines correspond to the frozen-in magnetic field, blue solid lines depict dependences obtained taking into account MAD in the presence of dust, red lines are the analytical solutions. Black dashed line on the left panel shows profile obtained using dependence $B \propto \rho^{1/2}$ describing magnetostatic contraction. Labels denote analytical slopes.}
\label{Fig:3}
\end{figure*}

Calculated radial profiles of vertical magnetic field component in the accretion disk of solar mass star are plotted in the panel (a) of figure \ref{Fig:3}. Panel (b) of the figure \ref{Fig:3} shows corresponding plasma parameter profiles in the midplane of the accretion disk,
\begin{equation}
	\beta_z = \frac{8\pi\rho V_s^2}{B_z^2}.
\end{equation}

Figure \ref{Fig:3}(a) shows that MAD leads to reduction of $B_z$ by order of magnitude comparing to the frozen-in field at $r>0.5$ AU. Plasma parameter corresponding to the frozen-in magnetic field depends on distance as $\beta_z \propto r^{-3/4}$ and it equals $\sim 10^3$ at $r=3$ AU (figure \ref{Fig:3}(b)). Dependence is more steep, $\beta_z \propto r^{-3/2}$, and $\beta_z \sim 10^4-10^5$ at $r=3$ AU, in the case of efficient MAD. Plasma parameter $\beta_z \approx 10-10^2$ in the outer regions of the accretion disk.

Comparison of frozen-in and MAD-profiles of $B_z$ allows to conclude that region of effective MAD, acting in $r$-direction, occupy significant part of the accretion disk from ionization minimum up to accretion disk outer boundary in the case of recombinations on dust grains. It should be noted that the profile of frozen-in $B_z$ is very similar to the magnetostatic profile $B \propto \rho^{1/2}$ that we used in the previous work \citep{dud13a}.

\begin{table*}
\small
\caption{Stellar magnetic field and magnetic field of accretion disk of solar mass star at typical distances.}
\centering
\begin{tabular}{lcccc}
\hline 
$m$ & $B_{s}(r_{\lti{in}})$, Gs & $B_z/B_z^{\lti{mad}}(r_{\lti{in}})$, Gs & $B_z/B_z^{\lti{mad}}(3\,\mbox{AU})$, Gs & $B_z/B_z^{\lti{mad}}(r_{\lti{out}})$, Gs \\ 
(1) & (2) & (3) & (4) & (5) \\ 
\hline 
$0.5$ & $2.6\,(0.085\,\mathrm{AU})$ & $0.6/0.6$ & $0.026/0.015$ & $0.0012/0.0012\,(70\,\mathrm{AU})$ \\ 
$1$ &  $11.5\,(0.052\,\mathrm{AU})$ & $2.4/2.4$ & $0.064/0.011$ & $0.0014/0.0011\,(140\,\mathrm{AU})$ \\ 
$1.5$ & $27.5\,(0.039\,\mathrm{AU})$ & $5.4/5.4$ & $0.11/0.013$ & $0.0015/0.0013\,(210\,\mathrm{AU})$ \\ 
$2$ &  $50.9\,(0.032\,\mathrm{AU})$ & $-/-$ & $0.13/0.020$ & $0.0013/0.0013\,(270\,\mathrm{AU})$ \\ 
\hline 
\end{tabular} 
\label{Tab:2}
\end{table*}

Table \ref{Tab:2} shows stellar mass dependence of $B_z$ strength at three typical distances. Non-dimensional stellar mass is shown in the first column. Stellar magnetic field at the inner boundary of the accretion disk, $B_s(r_{\lti{in}})$, is shown in the second column. Slash in the columns 3-5 separates values of frozen-in $B_z$ and that calculated taking into account MAD, $B_z^{\lti{mad}}$. Strengths of $B_z$ at the inner boundary of accretion disk (column 3), at 3 AU (column 4) and at the outer boundary $r_{\lti{out}}$ of the accretion disk (column 5) are shown. Corresponding values of $r_{\lti{in}}$ and $r_{\lti{out}}$ are presented in the brackets in columns 2 and 5.  Dash in the last row, column 3 means absence of solution. This can be explained in the following way.  Midplane temperature rises above $3000$ K at the distances $r<0.1$ AU for $m\geq 2$ and $\dot{M}\geq 4\times10^{-8}\,M_{\odot}/\mbox{yr}$. Non-monotonous behaviour of opacity $\kappa(T)$ at temperatures $T>3000$ K leads probably to thermal instability. \cite{belllin94} also pointed out that inner regions of the accretion disk cannot be correctly investigated in stationary approach.

Table \ref{Tab:2} shows that strength of frozen-in vertical magnetic field increases with stellar mass. Accretion disks of more massive stars are denser. Magnetic field increases with density growth according to (\ref{Eq:BzFrozen}, \ref{Eq:BzMADq}) and therefore amplification of the magnetic field is stronger in accretion disks of massive stars comparing to low-mass stars. Strength of $B_z$ is $(0.6-5)$ Gs at the accretion disk inner boundary which is several times lower than corresponding strength of stellar magnetic field $B_{s}(r_{\lti{in}})=(3-30)$ Gs. MAD is inefficient in this region according to figure \ref{Fig:3}. Strength of frozen-in $B_z$ at the distance $r=3$ AU is $(0.03-0.13)$ Gs depending on stellar mass. Efficient MAD leads to nearly one order of magnitude decrease of $B_z$, so that $B_z^{\lti{mad}}=(11-20)$ mGs at $r=3$ AU depending on the stellar mass. MAD is inefficient near the accretion disk outer boundary and the ${B_z(r_{\lti{out}}) \sim (1.1-1.5)\,\mathrm{mGs}}$.

\subsubsection{Magnetic field geometry}
\label{Sec:MFgeom}

We discuss geometry of the magnetic field calculated at the fiducial accretion disk and ionization rate parameters in this section. Effects of dust grain sizes, cosmic rays ionization rate and X-Ray luminosity on the magnetic field strength and geometry are considered.
\indexspace
{\it A. Dependence on dust parameters.}

\begin{figure*}[!htb]
\includegraphics[width=0.98\textwidth, trim=0cm 0cm 0cm 0cm]{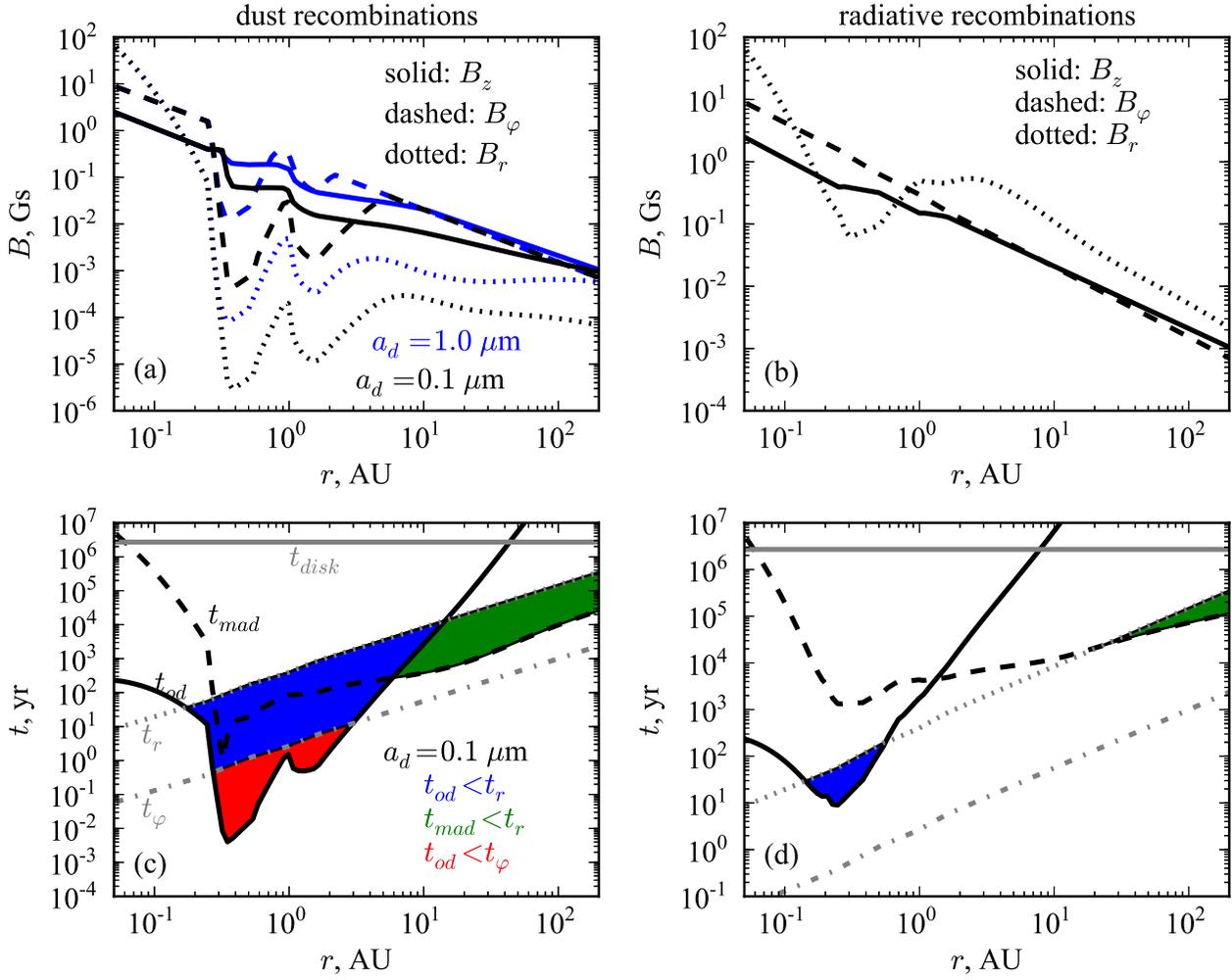}
\caption{{\it Panel a}: radial profiles of the magnetic field components in the accretion disk of solar mass star (dots: $B_r$, dashes: $B_{\varphi}$, solid line: $B_z$) calculated at height $z=H$ for recombinations on dust grains (black lines: $a_d=0.1\,\mu$m, blue lines: $a_d=1\,\mu$m). {\it Panel b}: The same as panel a, but for radiation recombinations. {\it Panel c}: dependences of the typical time scales on distance $r$ (gray solid line: accretion disk life time $t_{disk}$, solid black line: OD time scale $t_{od}$, dashed black line: MAD time scale $t_{mad}$, dotted gray line: time scale of $B_r$ generation $t_r$, dot-dashed gray line: time scale of $B_{\varphi}$ generation $t_{\varphi}$) calculated for recombinations on dust grains with $a_d=0.1\,\mu$m. {\it Panel d}: same as panel c, but for radiative recombinations.}
\label{Fig:4}
\end{figure*}

We plot magnetic field components versus the radial distance calculated at the different recombinations mechanisms and dust grains size in the top panels (a and b) of the figure \ref{Fig:4}. Radial profiles of the magnetic field components are calculated at the height $z=H$.

In figures \ref{Fig:4}(c, d) we show dependences of typical magnetic diffusion and dynamical times on $r$. All dependences correspond to the height $z=H$. Accretion disk lifetime is defined as $t_{disk} = M_{disk}/\dot M$. Time for OD $t_{od}=H/\eta_{od}$ and for MAD $t_{mad}=H/\eta_{mad}$, since magnetic diffusion acts in $z$ direction in the considered case. Azimuthal component of the magnetic field generation time $t_{\varphi}$ is estimated according to (\ref{Eq:Bphi_time}), and time of the $B_r$ generation is defined as $t_r = H / V_r$. It follows from equations (\ref{Eq:BrBphi}) that $t_{r}/t_{\varphi} \simeq 3/(2\alpha)$ so $t_r$ is approximately in $\alpha^{-1}$ times larger than $t_{\varphi}$. We assume that magnetic diffusion is effective if diffusion time is less than generation time and accretion disk lifetime. Figure \ref{Fig:4}(c) depicts the profiles obtained for the case of recombinations on dust grains with $a_d=0.1\,\mu$m, figure \ref{Fig:4}(d) -- for the radiative recombinations case. 

Figures \ref{Fig:4}(a, b) show that fossil magnetic field is amplified in $10^2-10^5$ times during accretion disk formation and evolution comparing to the magnetic field of protostellar clouds cores, $B_0\simeq 10^{-5}$ Gs. Amplification is stronger in the inner regions of the accretion disk where accretion velocity and density are larger. 
	
Figures \ref{Fig:4}(a, c) show that there are three regions in accretion disk with various magnetic field geometry in the case of recombinations on dust grains. 

{\it a)} OD and MAD are inefficient in the inner thermally ionized regions: $t_{od},\,t_{mad}>t_{r},\,t_{\varphi}$ at $r<0.5$ AU. Evolution of the magnetic field probably is non-stationary in this region. We assume that strength of the azimuthal component of magnetic field $B_{\varphi} \simeq 10$ Gs is limited by buoyancy. Magnetic field has the quasi-azimuthal geometry $B_{\varphi}>B_z,\,B_r$  in this region. 

{\it b)} Ohmic diffusion prevents $B_r$ and $B_{\varphi}$ generation in the region of lowest ionization fraction (``dead'' zone, see next section): $t_{od}<t_{r},\,t_{\varphi}$ at $r\in [0.5,\,3]$ AU. Magnetic field geometry is quasi-poloidal in this region, $B_z \gg B_{\varphi} \gg B_r$.

{\it c)} Magnetic ambipolar diffusion prevents the generation of $B_r$ in the outer regions of the accretion disk, $t_{mad}<t_r$ at $r > 10-20$ AU. Figure \ref{Fig:4}(c) shows that $t_{mad}>t_{\varphi}$ in this region, i.e. $B_{\varphi}$ is efficiently generated, $B_{\varphi}\sim B_z$. Magnetic field is quasi-azimuthal in this region, $B_{\varphi}\simeq B_z>B_{r}$. 

Figure \ref{Fig:4}(a) shows that intensity of magnetic field at $r>0.3$ AU is nearly order of magnitude higher in the case $a_d=1\,\mu$m comparing to the case $a_d=0.1\,\mu$m. This is explained by the fact that ionization fraction is higher in case $a_d=1\,\mu$m (see the figure \ref{Fig:2}(a)) and magnetic diffusion is less efficient. Geometry of the magnetic field is also quasi-poloidal at $r\in [0.5,\,1-2]$ AU for $a_d=1\,\mu$m. $B_r$ is comparable with the strength of $B_z$ at larger distances for $a_d=1\,\mu$m, since the ionization fraction is higher in this case comparing to the case $a_d=0.1\,\mu$m, and hence MAD of $B_r$ is less efficient. Magnetic field is quasi-radial near the accretion disk outer boundary in the case of large dust grains, $a_d\geq 1\,\mu$m.

In the pure radiative recombinations case, magnetic field is coupled to the matter: figure \ref{Fig:4}(d) shows that MAD is inefficient throughout the disk, $t_{mad}>t_r,\,t_{\varphi}$, and OD prevents $B_{r}$ generation only in the narrow distance range: $t_{od}<t_r$ at $0.1\,\mbox{AU}<r<1$ AU. Strength of the azimuthal magnetic field component is limited by buoyancy in this case, and $B_{\varphi} \simeq B_z$. Magnetic field is dynamically important under such circumstances. OD and MAD of $B_r$ are inefficient at $r>1$ AU. In this case, quasi-radial geometry of the magnetic field $B_r \gtrsim B_z\simeq B_{\varphi}$ may be responsible for generation of magneto-centrifugal winds in accordance with the criterion of \cite{blandford82}. We expect non-stationary behaviour of the magnetic field in this case.

\indexspace
{\it B. Dependence on ionization rates.}

\begin{figure*}[!htb]
\includegraphics[width=0.98\textwidth, trim=0cm 0cm 0cm 0cm]{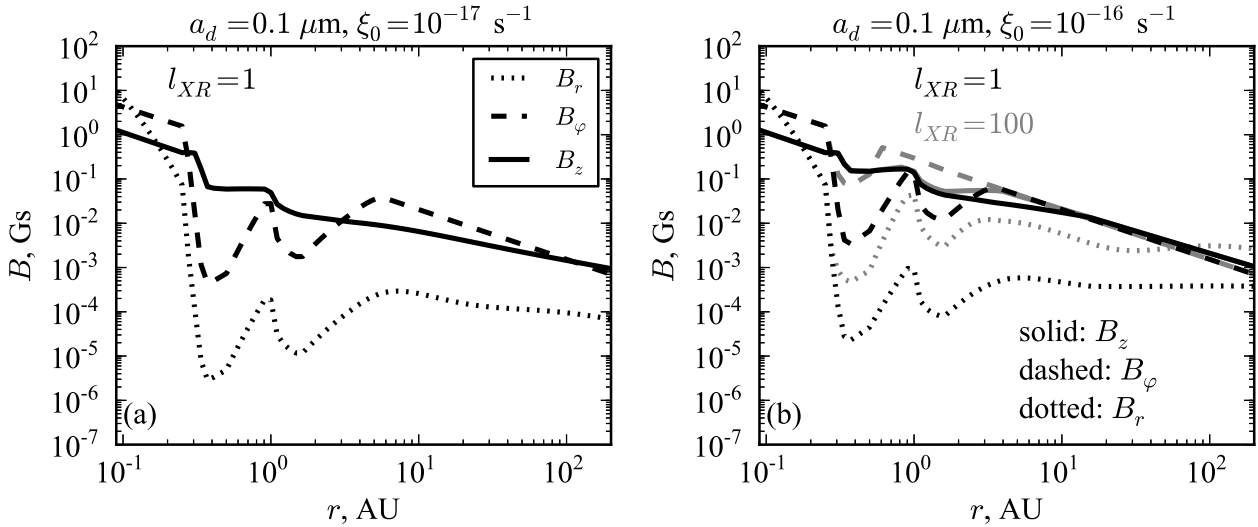}
\caption{Radial profiles of the magnetic field components (dots: $B_r$, dashes: $B_{\varphi}$, solid line: $B_z$) in the accretion disk of solar mass star at different ionization rates. {\it Left}: $\xi_{0}=10^{-17}\,\mbox{s}^{-1}$ and $L_{\lti{XR}}=10^{30}\,\mbox{erg}\,\mbox{s}^{-1}$. {\it Right}: $\xi_{0}=10^{-16}\,\mbox{s}^{-1}$ and $L_{\lti{XR}}=10^{30}\,\mbox{erg}\,\mbox{s}^{-1}$ (black lines), $\xi_{0}=10^{-16}\,\mbox{s}^{-1}$ and $L_{\lti{XR}}=10^{32}\,\mbox{erg}\,\mbox{s}^{-1}$ (blue lines).}
\label{Fig:5}
\end{figure*}

Figure \ref{Fig:5} shows dependence of magnetic field on ionization rates parameters, $\xi_0$ and $L_{XR}$. In the panel (a), we plot radial profiles of magnetic field components for fiducial ionization parameters. In figure \ref{Fig:5}(b) we vary cosmic rays ionization rate and X-ray luminosity in order to investigate their influence on the magnetic field geometry.

Increase of cosmic rays ionization rate from $10^{-17}\,\mbox{s}^{-1}$ to $10^{-16}\,\mbox{s}^{-1}$ leads to increase of the ionization fraction in $\sim$10 times (see figure \ref{Fig:2}). The MAD is less efficient in the case $10^{-16}\,\mbox{s}^{-1}$ and magnetic field strength is larger by order of magnitude comparing to the case $\xi_0=10^{-17}\,\mbox{s}^{-1}$, as figure \ref{Fig:5}(b) shows. Magnetic field is quasi-radial near the accretion disk outer edge under such circumstances, $B_r\simeq B_z\simeq B_{\varphi}$, apart from the case $\xi_0=10^{-17}\,\mbox{s}^{-1}$ when magnetic field is quasi-azimuthal $B_r < B_z\simeq B_{\varphi}$ (as in the figure \ref{Fig:5}(a)).

Figure \ref{Fig:5}(b) shows that growth of X-Rays luminosity by factor 100 leads to increase of $B_r$ by a factor 5-10 at distances $1-100$ AU. Vertical and azimuthal magnetic field components are frozen-in at distances $r>(3-4)$ AU. Therefore, magnetic field is quasi-radial, $B_r\simeq B_z\simeq B_{\varphi}$, in the outer regions of accretion disk in the case of high X-rays luminosity, $L_{\lti{XR}}> 10^{30}\,\mbox{erg}\,\mbox{s}^{-1}$.

\indexspace
{\it C. Dependence on accretion rate.}

\begin{figure}[!htb]
\includegraphics[width=0.48\textwidth, trim=0cm 0cm 0cm 0cm]{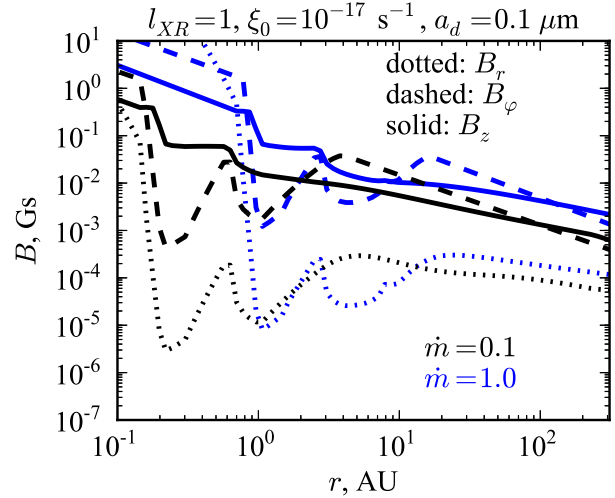}
\caption{Radial profiles of the magnetic field components (dots: $B_r$, dashes: $B_{\varphi}$, solid line: $B_z$) in the accretion disk of solar mass star calculated for different accretion rates (black lines: $\dot{M}=10^{-8}\,M_{\odot}/\mbox{yr}$, blue lines: $\dot{M}=10^{-7}\,M_{\odot}/\mbox{yr}$). }
\label{Fig:6}
\end{figure}

Figure~\ref{Fig:6} shows dependence of magnetic field components on the accretion rate.
We carry out calculations for $\dot{M}=10^{-8}\,M_{\odot}/\mbox{yr}$ and $\dot{M}=10^{-7}\,M_{\odot}/\mbox{yr}$. Smaller accretion rates correspond to more advanced stages of accretion disk evolution. Figure~\ref{Fig:6} shows that the magnetic field generally decreases as the accretion rate decreases. Density and temperature are smaller and generation of magnetic field is less efficient in accretion disks with smaller accretion rates. Mass accretion rate variation influences the strength of the magnetic field, while its geometry preserves, as figure \ref{Fig:6} and analytical solution (\ref{Eq:BzApprVisc}-\ref{Eq:BphiMADApprVisc}, \ref{Eq:BzApprIrr}-\ref{Eq:BphiMADApprIrr}) show.

\section{Dead zones}
\label{Sec:DZ}

 We treat ``dead'' zones as the regions of low ionization fraction and/or efficient magnetic diffusion. Magnetic diffusion suppresses MRI and hence MHD turbulence generation. Several different criteria of MRI dumping by magnetic diffusion were proposed: using MRI critical wavelength, magnetic Reynolds number and Elsasser number (see \cite{mohanty13}). Some uncertainty exists in definition of the critical magnetic Reynolds number. We adopt that MRI is suppressed if critical wavelength $\lambda_{\lti{cr}}$ exceeds scale height at given point of the disk. Critical wavelength in case of efficient magnetic diffusion is \citep[e.g.][]{kunz04}
\begin{equation}
	\lambda_{\lti{cr}} = \frac{2\pi}{\sqrt{3}}\frac{\eta}{V_{az}},\label{Eq:DZ}
\end{equation}
where Alfven velocity $V_{az} = B_z / \sqrt{4\pi\rho}$. We define OD-``dead'' zone as the region where OD suppresses MRI, and MAD-``dead'' zone as the region where MAD suppresses MRI, according to criterion $\lambda_{\lti{cr}} > H$. ``Dead'' zone boundary is determined as the locus $\lambda_{\lti{cr}} / H = 1$. Surface density of the active layer is calculated as $\Sigma_{\lti{al}}=\int\limits_{z_{\lti{dz}}}^\infty \rho dz$ where $z_{\lti{dz}}$ is the $z$-coordinate of the upper boundary of the ``dead'' zone.

\begin{figure*}[!htb]
\includegraphics[width=0.98\textwidth, trim=0cm 0cm 0cm 0cm]{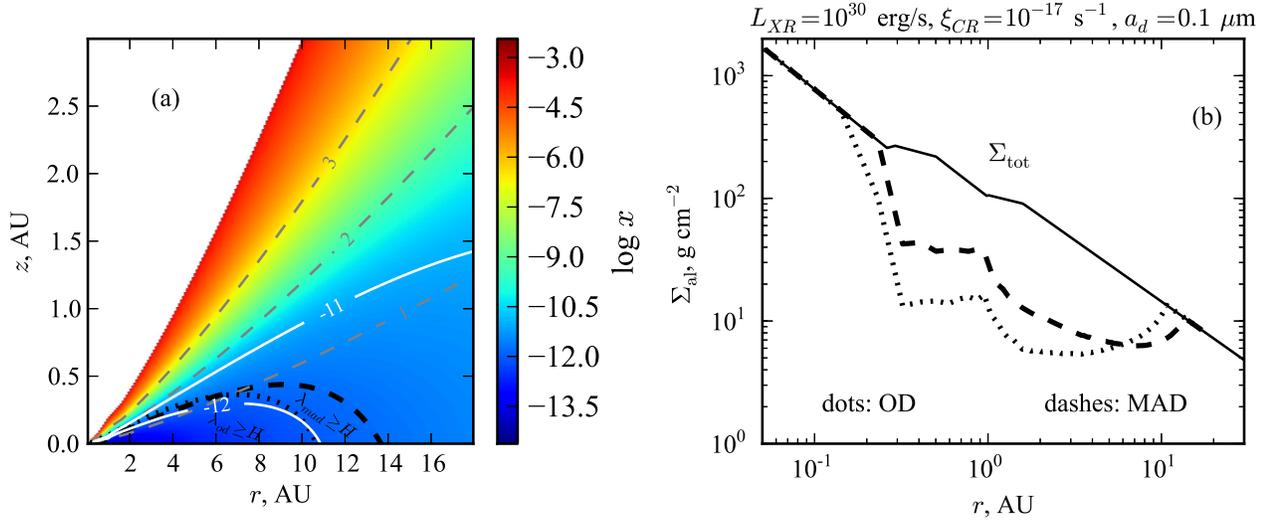}
\caption{``Dead'' zone structure in the accretion disk. {\it Left}: Two-dimensional ionization fraction distribution (colour map and white contours). Gray dashed lines depict accretion disk scale heights. Black lines bound OD-``dead'' zone (dots), $\lambda_{od}\geq H$,  and MAD-``dead'' zone (dashes), $\lambda_{mad}\geq H$. {\it Right}: Surface density of the active layers (dots: OD, dashes: MAD) and total surface density (solid).}
\label{Fig:7}
\end{figure*}

Figure \ref{Fig:7} shows the structure of the ``dead'' zone in the accretion disk of T Tauri star at the fiducial parameters. In figure \ref{Fig:7}(a), we plot ionization fraction distribution in $r-z$ plane and boundaries of the ``dead'' zone.
Surface density profiles of the active layers are shown in the figure \ref{Fig:7}(b).

Ionization fraction is the main parameter determining efficiency of OD and MAD. Boundary of the ``dead'' zone approximately coincides with the isosurface of critical ionization fraction \citep{gammie96}. Since ionization fraction has minimum at $r\sim 0.3$ AU (see section \ref{Sect:Ion}), ``dead'' zone has inner and outer boundaries. We denote coordinates of these boundaries as $r_{\lti{in}}^{\lti{dz}}$ and $r_{\lti{out}}^{\lti{dz}}$, respectively.  Surface density of the active layer is equal to the total accretion disk surface density at $r<r_{\lti{in}}^{\lti{dz}}$ and $r>r_{\lti{out}}^{\lti{dz}}$. Inner boundary of the ``dead'' zone is situated near the accretion disk inner boundary and it is not seen in the figure \ref{Fig:7}(a), where linear scale for distance $r$ is used. Figure \ref{Fig:7}(b) shows that OD-``dead'' zone is situated at distances from $\sim 0.2$ AU to $\sim 10$ AU. Ionization fraction $x \sim 10^{-12}$ at its boundary, as figure \ref{Fig:7}(a) shows.

MAD-``dead'' zone is located in the region from $\sim 0.3$ AU to $\sim 14$ AU. Hence, ``dead'' zone outer boundary is determined by MAD at the adopted parameters. Surface density of the active layer is lower than cosmic rays attenuation length. Although cosmic rays penetrate almost to the midplane, ionization fraction is small up to height $z \simeq H$ because of efficient recombinations on the dust grains. Ionization fraction exceeds boundary value $10^{-13}-10^{-12}$ at the high altitudes $z\simeq H$, according to the figure \ref{Fig:7}(a).

\subsection{Dependence on ionization rates}

We calculate the surface density of active layer for different ionization rates and dust grain sizes. We vary one parameter in each calculation while fiducial values of the rest parameters are used. 

As it was found in the section \ref{Sect:Ion}, growth of dust grain size in 10 times leads to increase of the ionization fraction in 10 times. Consequently, size of the region of the efficient magnetic diffusion decreases. We found that the outer boundary of the ``dead'' zone is situated closer to the star, $r_{\lti{out}}^{\lti{dz}}\simeq 5$ AU in the case $a_d=1\,\mu$m comparing to the case $a_d=0.1\,\mu$m for which $r_{\lti{out}}^{\lti{dz}}\simeq 14$ AU. Surface density of the active layer is several times larger in the disk with large dust grains ($a_d=1\,\mu$m) in comparison to the disk with the small dust grains ($a_d=0.1\,\mu$m).

Ionization fraction increases by several orders of magnitude in the case of radiative recombinations. Our calculations show that OD-``dead'' zone is situated between $r_{\lti{in}}^{\lti{dz}}\simeq 0.3$ AU and $r_{\lti{out}}^{\lti{dz}}\simeq 0.5$ AU in this case. Surface density of the active layer $>100\,\mbox{g}\,\mbox{cm}^{-2}$ here. MAD is inefficient in the case of radiative recombinations (see section \ref{Sec:MFgeom}) and it does not lead to ``dead'' zone formation.

Outer boundary of the ``dead'' zone is situated at $r_{\lti{out}}^{\lti{dz}}\simeq 4-6$ AU in the case when $\xi_{0}=10^{-16}\,\mbox{s}^{-1}$ that is in 10 times larger then the fiducial value. Surface density of the active layers is almost the same as for $\xi_{0}=10^{-17}\,\mbox{s}^{-1}$. Increasing of the X-ray luminosity by factor 100 also leads to reduction of the ``dead'' zone sizes. Outer boundary of the ``dead'' zone is placed at $r_{\lti{out}}^{\lti{dz}}\simeq 7$ AU in the case of high X-ray luminosity, $L_{XR}=10^{32}\,\mbox{erg}\,\mbox{s}^{-1}$. Surface density of active layers $\Sigma_{\lti{al}}\gtrsim 30\,\mbox{g}\,\mbox{cm}^{-2}$ under such circumstances.

Our calculations show that varying ionization parameters affect only outer boundary of the ``dead'' zone and surface density of the active layers. OD determines the inner boundary of the ``dead'' zone. This boundary is situated in the region of metals thermal ionization at the distance $r\sim 0.2$ AU for solar mass star. Assuming that thermal  ionization begins at $T= 1000$ K, we derive following estimation for $r_{\lti{in}}^{\lti{dz}}$ from the analytical solution (\ref{Eq:TcAppr})
\begin{equation}
	r_{\lti{in}}^{\lti{dz}} = 0.28\,\alpha_{0.01}^{-2/9}\dot{m}^{4/9}m^{1/3}\,\mbox{AU},\label{Eq:DZin}
\end{equation}
Estimation (\ref{Eq:DZin}) agrees with the approximation of \citet{kretke09}.

\subsection{Dependence on stellar masses}

Table \ref{Tab:3} shows the accretion disk and ``dead'' zone characteristics calculated for different stellar masses in the case of recombinations on the dust grains. Fiducial ionization parameters are used. Non-dimensional stellar mass and luminosity are shown in the columns 1 and 2, respectively.  Positions of the accretion disk inner and outer boundaries, masses of the accretion disks are listed in columns 3-5. Positions of ``snow'' line ($r_{\lti{sl}}$, distance where ice evaporation begins, at $T = 150$ K) at the midplane of the accretion disk are shown in the column 6. Positions of the ``dead'' zone inner $r_{\lti{in}}^{\lti{dz}}$ and outer boundaries $r_{\lti{out}}^{\lti{dz}}$, mass contained in the ``dead'' zone $M_{\lti{dz}}$ (in Jupiter masses) are listed in the columns 7-12. Critical ionization fraction at the ``dead'' zone boundaries $x_{cr}$ is shown in the columns 13 and 14. Columns 7, 9, 11 and 13 show characteristics of the OD-``dead'' zone, columns 8, 10, 12, 14 -- of the MAD-``dead'' zone. 

\begin{table*}
\small
\caption{Accretion disk and ``dead'' zone parameters}
\centering
\begin{tabular}{rrrrrrrrrrrrrr}
\hline 
$m$ & $L_{\star}$ & $r_{\lti{in}}$ & $r_{\lti{out}}$ & $M_{\lti{disk}}$ & $r_{\lti{sl}}$ & \multicolumn{2}{c}{$r_{\lti{in}}^{\lti{dz}}/$AU} & \multicolumn{2}{c}{$r_{\lti{out}}^{\lti{dz}}/$AU} & \multicolumn{2}{c}{$M_{\lti{dz}} / M_{\lti{J}}$} &  \multicolumn{2}{c}{$x_{cr}$}\\ 
\hline
&  ($L_{\odot}$) &  (AU) & (AU) &  ($M_{\odot}$) & (AU) & OD & MAD & OD & MAD & OD & MAD & OD & MAD\\ 
(1) &  (2) &  (3) & (4) &  (5) & (6) & (7) & (8) & (9) & (10) & (11) & (12) & (13) & (14) \\ 
\hline
0.5	& 0.5	& 0.085	& 70	& 0.003 & 2.5	& 0.11	& 0.14	& 3.1	& 2.8	& 0.04	& 0.02 & $2.7\times 10^{-12}$ & $2.2\times 10^{-12}$\\ 
1	& 1		& 0.052	& 140	& 0.027	& 3.6	& 0.26	& 0.32	& 10.7	& 13.7	& 0.8	& 1.0 & $9.2\times 10^{-13}$ & $4.6\times 10^{-12}$\\ 
1.5	& 1.5	& 0.039	& 210	& 0.1	& 4.3	& 0.43	& 0.55	& 16.3	& 19.0	& 3.6	& 4.1 & $5.8\times 10^{-13}$ & $8.4\times 10^{-13}$\\ 
2	& 10	& 0.032	& 270	& 0.17	& 11.0	& 0.60	& 0.75	& 17.2	& 21.4	& 5.0	& 5.9 & $5.3\times 10^{-13}$ & $9.1\times 10^{-13}$\\ 
\hline 
\end{tabular}
\label{Tab:3}
\end{table*}

Table \ref{Tab:3} shows that the inner boundary of the ``dead'' zone is situated at $r=0.1-0.6$ AU depending on stellar mass which agrees with analytical estimation (\ref{Eq:DZin}). Comparison of the columns 9 and 10 shows that MAD-``dead'' zone extent is by 3-4 AU larger than OD-``dead'' zone extent for $m>0.5$, $r_{\lti{out}}^{\lti{dz}} = 3-21$ AU depending on stellar mass. This result agrees with results of section \ref{Sec:MFgeom} indicating that MAD is the main magnetic diffusion mechanism in the outer accretion disk regions.

Ionization fraction is lower and magnetic diffusion is more efficient in the accretion disks of massive stars, since these disks are denser. Consequently, size and mass of the ``dead'' zone increase with stellar mass. In the accretion disk of solar mass star $r_{\lti{out}}^{\lti{dz}}\sim 14$ AU and $M_{\lti{dz}}=1.0\,M_{\lti{J}}$. These values are comparable with corresponding protosolar nebula parameters \citep{weidenschiling77}. ``Snow'' line lies inside ``dead'' zone for all stellar masses.

We adopted standard dust-to-gas mass fraction $Y_{\lti{g}}=0.01$ in the calculations. Hence, total mass of solid particles inside ``dead'' zone is 0.13, 3.2, 13.0 and 19.0 $M_{\oplus}$ for stars with masses 0.5, 1, 1.5 and 2 $M_{\odot}$, respectively.  Mass of solid material contained in the ``dead'' zones is more than $3\,M_{\oplus}$ in accretion disks of stars with $M\geq 1\,M_{\odot}$. This mass of dusty material is sufficient for formation of several embryos of the Earth type planets in this region. Formation of solid cores with mass $M_{\lti{c}}>10\,M_{\oplus}$ can trigger gas giant planet formation by core accretion \citep{armitage_book}.

Ionization fraction $x_{cr} \sim 10^{-13}-10^{-12}$ at the boundary of the ``dead'' zone. Boundary ionization fraction $x_{cr}$ is several times larger for MAD-``dead'' zone than for OD-``dead'' zone.

\section{Discussion and conclusions}
\label{Sec:Discussion}

We elaborate kinematic MHD model of accretion disks of young stars. The model is based on Shakura and Sunyaev approximations. Magnetic field is calculated from stationary induction equation taking into account the ohmic diffusion (OD), magnetic ambipolar diffusion (MAD) and buoyancy. Ionization fraction is determined taking into consideration thermal ionization of metals and shock ionization by cosmic rays, X-rays and radioactive elements. Recombinations on dust grains, radiative recombinations and dust evaporation are included in the ionization model.

Accretion disk sizes and masses calculated in the frame of our model agree with observations. Inner boundary of the accretion disk is placed at the distances $r=0.03-0.09$ AU (several stellar radii), outer boundary is placed at $r=70-270$ AU depending on stellar mass in the interval 0.5-2 $M_{\odot}$. Corresponding masses of accretion disks equal 0.003-0.17 of solar mass. We derive an analytical solution of the model equations in the case when dependence of ionization fraction on gas density is the power-law function.

Attenuation of cosmic rays and X-rays leads to decrease of ionization fraction to extremely low values $\sim 10^{-16}-10^{-14}$ in the inner dense cold regions of accretion disk in the case of recombinations on the dust grains. Magnetic diffusion is efficient in this region of lowest ionization fraction (``dead'' zone). OD develops in the ``dead'' zone mainly where ionization fraction is minimal. MAD operates in the outer regions of the accretion disk. Magnetic field is frozen-in near the inner boundary of the accretion disk, where thermal ionization of metals with low ionization potential takes place.
	
The fossil magnetic field of accretion disks of young stars has complex geometry. Initially uniform magnetic field of parent protostellar clouds transforms in accretion disk. All three components of the magnetic field are not zero. Differential rotation generates azimuthal component of the magnetic field. Accretion generates radial component of the magnetic field. We derive the following estimation for geometrically thin disk: $B_r\simeq 2/3\,\alpha B_{\varphi}$, were $\alpha$ is Shakura and Sunyaev parameter. Hence, radial component $B_r$ typically is much smaller than the azimuthal component $B_{\varphi}$ while their dependence on radial distance is the same.

The magnetic field is qiasi-poloidal in the dusty ``dead'' zones. Estimations of the characteristic times of magnetic diffusion show that times of OD and MAD are less than lifetime of accretion disk in ``dead'' zones. Therefore, magnetic field geometry remains purely poloidal in this region. Strength of the vertical magnetic field is $B_z\simeq (10-20)$ mGs at $r=3$ AU depending on stellar mass, $M=0.5-2\,M_{\odot}$. Sedimentation of charged dust grains is possible in the ``dead'' zones along magnetic field lines.

Magnetic field is coupled to the matter in the inner regions, $r\lesssim 0.3$ AU. Magnetic field is quazi-azimuthal in this region. Strength of magnetic field is comparable with stellar magnetic field strength at the accretion disk inner boundary, which equals 5-30 Gs depending on stellar mass. This means, that accretion disk inner boundary must be determined taking into account pressure of the accretion disk magnetic field. Interaction of the stellar magnetic field and accretion disk magnetic field near the accretion disk inner boundary may lead to current sheet formation that can become an additional source of X-ray activity of stellar magnetospheres.

Magnetic field is quazi-azimuthal or quasi-radial in the outer regions of the accretion disk depending on intensity of ionization mechanisms. Growth of dust grains to $a_d\geq 1\,\mu$m, or increase of cosmic rays ionization rate to $10^{-16}\,\mathrm{s}^{-1}$, or increase of X-rays luminosity to $10^{32}\,\mathrm{erg}\,\mathrm{s}^{-1}$ are necessary to form quasi-radial magnetic field. The quasi-radial geometry of the magnetic field may be responsible for generation of magneto-centrifugal winds in the outer regions of the accretion disks owing to the criterion of \cite{blandford82}.

Midplane ionization fraction does not fall below $10^{-11}$ in the case of radiative recombinations. Magnetic diffusion is inefficient and magnetic field is coupled to gas throughout the accretion disk. Vertical magnetic field is proportional to the surface density in this case. Our calculations show that $B_z(3\,\mbox{AU})=(0.03-0.13)$ Gs depending on stellar mass. Magnetic field is quasi-azimuthal, and $B_{\varphi}\simeq B_z$ at the distances less than several AU.  Magnetic field is quasi-radial and magnetically-driven outflows are possible in the outer regions of accretion disks.

Complexity of the fossil magnetic field geometry in accretion disks of young stars makes its investigation with Zeeman experiments and polarization measurements difficult. It requires resolution 0.1 AU at minimal distance 1 pc that corresponds to angular resolution $10^{-6}$ angular seconds. However, it is doubtful that, at present time, separation of Zeeman lines splitting from turbulent broadening of the lines in the accretion disk is possible. This is feasible for the ``dead'' zones only.

``Dead'' zones are very attractive for observations, both in terms of measurements of magnetic field strength, considering it is poloidal, and in terms of Earth type planet formation. Turbulence is weak and dust sedimentation is efficient here. Our calculations show that MAD determines the outer boundary of the ``dead'' zone for stars with $M>0.5\,M_{\odot}$. Outer boundary of the ``dead'' zone is placed at the distances 3-21 AU depending on the stellar mass in the case of recombinations on dust grains. 

In the case when dust is present in the disk, surface density  of active layers is $\sim 10\,\mathrm{g}\,\mathrm{cm}^{-2}$ at $r\sim 1$ AU for typical parameters, i.e. ``dead'' zone occupies significant part of accretion disk thickness. Angular momentum transport by MRI-induced turbulence is inefficient under such circumstances. Tension of large-scale magnetic field lines may be responsible for the angular momentum transport in this region.

Mass of solid material inside ``dead'' zones in accretion disks of stars with $M\geq 1\, M_{\odot}$ is more than 3 $M_{\oplus}$. This mass is sufficient for formation of several embryos of the Earth type planets in this region. Collisional formation and growth of planetesemals are the possible mechanisms of embryos formation \citep{safronov_book}.

Our work is the step forward in the investigations of magnetized accretion disks of young stars comparing to existed semi-analytical investigations. In order to calculate magnetic field strength and geometry, we incorporate induction equations into the accretion disk model. Adopted kinematic approximation allows us investigate magnetic field structure with the help of quite simple semi-analytical equations of the model. We investigate ohmic diffusion and magnetic ambipolar diffusion in detail without paying attention to the Hall effect that is believed to influence the evolution of MHD turbulence in accretion disks of young stars \citep[e.g.,][]{wardle12}. Turbulence in the accretion disks also must influence magnetic field dissipation and generation. We intend to investigate the Hall effect and turbulent diffusion influence on the fossil magnetic field strength and geometry in accretion disks in our next paper. In order to improve elaborated model, we need to take into account dynamical influence of magnetic field on the accretion disk structure.

\acknowledgments

We thank anonymous referee for some useful comments.

\nocite{*}
\bibliographystyle{spr-mp-nameyear-cnd}
\nopagebreak
\bibliography{dudorov_bib}

\end{document}